\documentclass{aa}  
\usepackage{graphicx}
\usepackage{txfonts}
\usepackage{natbib}
\usepackage[table]{xcolor}
\usepackage{hyperref}
\usepackage{psfrag}
\definecolor{gris25}{gray}{0.80}

\def\micron{\,$\mu$m}

\def\arcsec{"}

\def\aap{A\&A}

\def\apjl{ApJL}

\def\kms{\,${\rm km}\cdot{\rm s}^{-1}$\,}

\def\rsun{R$_{\odot}$\,}

\begin{document}
%


\authorrunning{F. Millour et al.}
\titlerunning{a binary engine fuelling HD\,87643's complex environment}

\title{A binary engine fuelling HD\,87643's complex circumstellar
  environment
  \thanks{
    Based on observations made with the ESO Very Large Telescope at
    Paranal Observatory under programs 076.D-0575, 077.D-0095,
    076.D-0141, 380.D-0340, and 280.C-5071, with the ESO 1.52-m
    and archival ESO data.
  }
}

\subtitle{determined using AMBER/VLTI imaging}

\author{
  F.~Millour\inst{1}
  \and
  O.~Chesneau\inst{2}
  \and
  M.~Borges Fernandes\inst{2}
  \and
  A.~Meilland\inst{1}
  \and\\
  G. Mars\inst{3}
  \and
  C. Benoist\inst{3}
  \and
  E.~Thi\'ebaut \inst{4}
  \and
  P.~Stee \inst{2}
  \and
  K.-H.~Hofmann\inst{1}
  \and
  F.~Baron \inst{5}
  \and
  J.~Young \inst{5}
  \and\\
  P.~Bendjoya\inst{2}
  \and
  A.~Carciofi \inst{6}
  \and
  A.~Domiciano~de~Souza \inst{2}
  \and
  T.~Driebe \inst{1}
  \and
  S.~Jankov \inst{2}
  \and
  P.~Kervella \inst{7}
  \and
  R.~G.~Petrov \inst{2}
  \and
  S.~Robbe-Dubois \inst{2}
  \and
  F.~Vakili \inst{2}
  \and
  L.B.F.M.~Waters \inst{8}
  \and
  G.~Weigelt \inst{1}
}

\offprints{
  F.~Millour\\
  \email{FMillour@mpifr-bonn.mpg.de}
}

\institute{
  Max-Planck Institut f\"ur Radioastronomie, Auf dem H\"ugel 69,
  53121, Bonn, Germany,
  \and
  UMR 6525 H. Fizeau, Univ. Nice Sophia Antipolis, CNRS, Observatoire
  de la C\^{o}te d'Azur, F-06108 Nice cedex 2, France 
  \and
  UMR 6202 Cassiop\'{e}e, Univ. Nice Sophia Antipolis, CNRS,
  Observatoire de la C\^{o}te d'Azur, BP 4229, F-06304 Nice, France
  \and
  UMR 5574 CRAL, Univ. Lyon 1, Obs. Lyon, CNRS, 9 avenue
  Charles Andr\'e, 69561 Saint Genis Laval cedex, France
  \and
  Astrophysics Group, Cavendish Laboratory, University of Cambridge,
  J.J. Thomson Avenue, Cambridge CB3 0HE, UK
  \and
  Instituto de Astronomia, Geof\'isica e Ci\^encias Atmosf\'ericas,
  Universidade de S\~ao Paulo, Rua do Mat\~ao 1226, Cidade
  Universit\'aria, 05508-900, S\~ao Paulo, SP, Brazil
  \and
  LESIA, Observatoire de Paris, CNRS UMR 8109, UPMC,
  Univ. Paris Diderot, 5 place Jules Janssen, 92195 Meudon Cedex, France
  \and
  Faculty of Science, Astronomical Institute `Anton
  Pannekoek', Kruislaan 403, 1098 SJ Amsterdam, The Netherlands
}



\abstract
{
  The star HD\,87643, exhibiting the ``B[e] phenomenon'', has one
  of the most extreme infrared excesses for this object
  class. It harbours a large amount of both hot and cold dust, and
  is surrounded by an extended reflection nebula.
}
{
  One of our major goals was to investigate the presence of a
  companion in HD87643. 
  In addition, the presence of close dusty material was
  tested through a combination of multi-wavelength high spatial
  resolution observations.
}
{
  We observed HD\,87643 with high spatial resolution techniques, using
  the near-IR AMBER/VLTI interferometer with baselines ranging from
  60\,m to 130\,m and the mid-IR MIDI/VLTI interferometer with baselines
  ranging from 25\,m to 65\,m. These observations are complemented by
  NACO/VLT adaptive-optics-corrected images in the K and L-bands, and
  ESO-2.2m optical Wide-Field Imager large-scale images in the B, V
  and R-bands,
}
{
  We report the direct detection of a companion to HD\,87643 by means
  of image synthesis using the AMBER/VLTI instrument. The presence
  of the companion is confirmed by the MIDI and NACO
  data, although with a lower confidence. The
  companion is separated by $\sim34$\,mas with a roughly
  north-south orientation. The period must be large (several tens
  of years) and hence the orbital parameters are not determined
  yet. Binarity with high eccentricity might be the key to
  interpreting the extreme characteristics of this system, namely
  a dusty circumstellar envelope around the primary, a compact
  dust nebulosity around the binary system and a complex extended
  nebula suggesting past violent ejections. 
}
{}

\keywords{
  Techniques: high angular resolution --
  Techniques: interferometric  --
  Stars: emission-line, Be  --
  Stars: mass-loss --
  Stars: individual (HD\,87643) --
  Stars: circumstellar matter}
\maketitle
%

\section*{Introduction}

Stars with the ``B{[}e{]} phenomenon'' are B-type stars with
strong Balmer emission lines, numerous permitted Fe\,{\sc ii}
lines, and forbidden O\,{\sc i} and Fe\,{\sc ii} lines in their
optical spectrum. In addition, these stars exhibit a strong near
and mid-IR excess due to circumstellar dust
\citep{1976A&A....47..293A, 1997ASPC..120..161C}. For brevity,
we henceforth refer to stars showing the B{[}e{]} phenomenon as
``B[e] stars''.  The class of B[e] stars is composed of several
sub-classes, including objects at different evolutionary
stages and of low, medium, and high mass \citep{1998A&A...340..117L}.
Distances toward them are
usually poorly known. As a consequence, the determination of the
evolutionary state of B[e] stars is often uncertain. 


Observations suggest that the wind of some of the most massive B[e] stars
\citep[the so-called supergiant B{[}e{]} stars or sgB{[}e{]}
in][]{1998A&A...340..117L} is composed of two
distinct components, as proposed by \citet{1985A&A...143..421Z}. The
first component is a wind of low density in the polar regions; the
second component is a wind of high density and low velocity located in
the equatorial region of the star. The thermal infrared excess is
supposed to be produced in the outer parts of the equatorial wind
where the temperature allows the formation of dust grains, but dust
might also survive much closer to the star in a dense and compact disc.

The formation of dusty environments around B[e] evolved objects is
a challenge, with many theoretical issues
\citep[see][]{2003A&A...405..165K}. A non-spherical, disc-like
environment may be the key to understanding dust formation in sgB[e]
stars. It could be attributed to the rapid rotation of these stars
\citep[e.g.][]{1991A&A...244L...5L, 1993ApJ...409..429B,
  2006A&A...456..151K, 2008A&A...478..543Z} but could also be
caused by companions, since some B[e] stars were found to be
binaries. Indeed, the number of companions detected around these
objects is steadily growing \citep{2004A&A...417..731M,
  2006ASPC..355..347M}, but claims that all B[e] stars are
binaries are still controversial \citep{2003A&A...408..257Z}.

The number of B[e] stars is rather limited, and many of them are
unclassified \citep[cited as unclB{[}e{]}
in][]{1998A&A...340..117L} because they exhibit properties usually
associated with young and evolved objects simultaneously.

Our group has undertaken a large observing campaign to investigate
these poorly studied objects (sgB[e] and unclB[e]) mostly using
optical interferometry with the Very Large Telescope Interferometer
\citep[hereafter VLTI, see][for an introduction to interferometric
observations of Be and B{[}e{]}
stars]{2005ASPC..337..211S}. Early results were published on the
sgB[e] star {CPD-57 2874} \citep{2007A&A...464...81D}.

HD\,87643 (Hen 3-365, MWC 198, IRAS~10028-5825)  appears to be a
special case among sgB[e] and unclB[e] stars. It is a B2[e]
star \citep{1998MNRAS.300..170O} that exhibits the largest infrared
excess among this class and appears to be embedded in a complex
nebula whose properties are reminiscent of the nebulae around LBVs
\citep{1972PASP...84..594V, 1981A&A....93..285S,
  1983A&A...117..359S}. HD\,87643 is considered a unclB[e] by
\citet{1998A&A...340..117L}: it was classified as sgB[e]
\citep[see the estimation of the bolometric luminosity
in][]{1988ApJ...324.1071M, 2001A&A...368..160C}, under the assumption
that it lies close to the Carina arm
\citep[2-2.5\,kpc][]{1998MNRAS.300..170O, 1998ASSL..233..145M,
  1988ApJ...324.1071M}, but one can also find
it classified as a Herbig star \citep{2000ApJS..129..399V,
  2006MNRAS.367..737B}. The $v \sin i$ of the central star is unknown
since no pure photospheric line is detected in the visible spectrum.
Clues to the disc-like geometry of the environment are provided by
the polarisation ellipse in the U-Q plane across H$\alpha$
\citep{1998MNRAS.300..170O}. The detected PA is about
15$^\circ$-20$^\circ$ \citep{1998A&AS..131..401Y,
  1998MNRAS.300..170O}.  
On the other hand, \citet{2006MNRAS.367..737B} claimed to have
detected an asymmetric outflow around HD87643 using
spectro-astrometry. 
However, they also note that ``HD\,87643 stands out in the complexity
of its spectro-astrometry'', compared with the numerous other Herbig
Be stars they observed.

HD\,87643 has displayed a long-term decline of the visual
brightness for the last 30\,yrs, from $V\approx8.5/8.8$ in 1980
\citep[][]{1998ASSL..233..145M}, to $V\approx9.3/9.4$ in early
2009 \citep{2009ASPC..403...52P}. This decline is superimposed
on shorter-term variations, such as a 0.5\,mag
decrease in $\approx$1\,month, followed by a 0.7\,mag increase
in $\approx$5 months, seen in the ASAS photometric survey
\citep{2009ASPC..403...52P} between JD~2453000 and
JD~2453180. \citet{1998A&AS..131..401Y} also noted that HD87643
shows Algol-like variability, the star showing variations ``on a
timescale of days'', and being ``bluer during brightness
minima''.

The totality of the ISO/SWS 2-45\micron\ mid-IR emission from
HD\,87643 seems concentrated within the smallest ISO aperture (14"
x 22"), and at 10\micron, the dusty environment is unresolved on a
1" scale \citep{1999PhDT.........3V}. The ISO/LWS spectrum also
continues from the shorter-wavelength SED without any jump, 
which is an additional sign that the mid-IR emission is
compact. The disc-like geometry is strongly supported by the
huge infrared excess exhibited by the source, the large absorption
of the central star flux, and the polarimetric
data. \citet{1999PhDT.........3V} proposed that the disc might be a
circumbinary disc.

In this paper we report new observations that bring a new
insight into this interesting object, proving the presence
of a companion, a resolved circum-primary dust envelope (most
likely a disc), and circumbinary material. 

The outline of the article is as follows: the observations and
data recorded are presented in Sect.~\ref{sect:obsanddataproc}, then
we present and discuss the main results from our observing campaign in
Sect.~\ref{sect:newfacts}, and give a global view of the system,
considering this new information, in Sect.~\ref{sect:discussion}.

\section{Observations and data processing}
\label{sect:obsanddataproc}

\subsection{AMBER/VLTI near-IR interferometry}

HD\,87643 was observed at the ESO/Paranal observatory with the
Astronomical Multi BEam Recombiner (AMBER), the
near-infrared instrument of the VLTI \citep{2007A&A...464....1P}. The
observations were carried out on February 18, 2006 in medium
spectral resolution ($R=1500$) and during a series of nights in March 2008
at low spectral resolution ($R=35$). AMBER uses three 8m
telescopes (Unit Telescopes, hereafter UT) or three
1.8m telescopes (Auxiliary Telescopes, hereafter
AT). The calibration stars used were HD\,109787, HD\,86440,
HD\,101531, HD\,63744 and $\epsilon$ Oph. Details of the observations can
be found in Table~\ref{tab:log_obs}.

\begin{table}[htbp]
  \begin{caption}
    {
      AMBER and MIDI observing logs.
    }
    \label{tab:log_obs}
  \end{caption}
  \centering
  \begin{tabular}{lccc}
    \hline
    & & \multicolumn{2}{c}{projected baseline} \\
    Date & Stations & Length & PA \\
    & & [meter] & [degrees] \\
    \hline
    \hline
    \multicolumn{4}{c}{AMBER 1 (UT)}\\
    \hline
    18/02/2006, 3h & UT1-3-4 & 95, 57, 129 & 22, 88, 46 \\
    18/02/2006, 8h & UT1-3-4 & 71, 62, 102 & 65, 145, 102\\ 
    \hline
    \multicolumn{4}{c}{AMBER 2 (AT)}\\
    \hline  
    01/03/2008, 2h & K0-G1-A0 & 83, 78, 128 & -168, -94, -132 \\ 
    05/03/2008, 3h & G1-D0-H0 & 63, 64, 60.0 & -59, 64, 3 \\ 
    06/03/2008, 5h & G1-D0-H0 & 69, 60, 58 & -37, 90, 19\\ 
    06/03/2008, 7h & G1-D0-H0 & 71, 52, 33 & -15, 117, 33\\ 
    10/03/2008, 3h & H0-G0-E0 & 31, 15.6, 46.8 & -102 \\ 
    11/03/2008, 0h & H0-G0-E0 & 32, 16, 47 & -147 \\ 
    12/03/2008, 0h & H0-G0-E0 & 32, 16, 48 & -145 \\ 
    \hline   
    \multicolumn{4}{c}{MIDI (AT)}\\
    \hline   
    26/02/2006, 4h & D0-G0 & 31.3 & 76.3  \\
    27/02/2006, 6h & A0-G0 & 57.1 & 101.3 \\
    01/03/2006, 1h & A0-G0 & 63.5 & 38.1  \\
    01/03/2006, 5h & A0-G0 & 61.1 & 85.1  \\
    19/04/2006, 4h & D0-G0 & 25.4 & 123.1 \\
    23/05/2006, 2h & A0-G0 & 51.7 & 120.0 \\
    25/05/2006, 1h & A0-G0 & 54.4 & 110.6 \\
    \hline
  \end{tabular}
\end{table}

The data were processed with the standard AMBER data reduction software
\citep[\texttt{amdlib} 2.1, see for instance][]{2007A&A...464...29T,
  2004SPIE.5491.1222M} plus a series of advanced scripts to calibrate
the data \citep{2008SPIE.7013E.132M}.
The AMBER DRS performs a fringe fitting instead
of Fourier transforms and computes interferometric data products such
as $V^2$ and closure phases \citep[for a review of the
interferometric data analysis, see:][]{2006NewAR.51...583H,
  2006NewAR.51...604M, 2008NewAR..52..177M}. 
We performed the reduction using standard selection criteria \citep[see for
instance appendix C in][]{2007A&A...464..107M} for the individual
exposures, rejecting 80\% of the data before averaging the
data products. The additional scripts 
allowed us to compute realistic error bars, including the
uncertainties on the diameters of the calibration stars, the instrument
atmosphere transfer function instabilities, and the fundamental
noise. The study of the transfer function provided by different
calibrators gives an estimate of absolute errors on $V^2$
between 0.05 and 0.1. The visibilities and closure phases are
shown in Fig.~\ref{fig:AMBER_V2}.

\subsection{MIDI/VLTI mid-IR interferometry}

The observations of HD\,87643 at the ESO/Paranal observatory using
the MID-Infrared instrument \citep[hereafter
MIDI,][]{2003SPIE.4838..893L, 2004A&A...423..537L} were carried
out from February until May 2006. MIDI is the mid-infrared
(N-band, 7.5-13.5\micron) two-telescope combiner of the VLTI,
operating like a classical Michelson interferometer. In our case,
only the ATs were used since HD\,87643 is bright in the mid-IR
(156\,Jy). We used a standard MIDI observing sequence, as described
by \citet{2007A&A...471..173R}. The source was observed in the
so-called High-Sens mode, implying that the photometry from each
individual telescope is performed subsequently to the recording of
the fringes. The low spectral resolution (R=30) provided by the
prism was used. The data consist of 7 visibility spectra and one
flux spectrum (PSF 1.2$\arcsec$ with ATs). The errors on the
visibilities, including the internal ones and those from the
calibrator diameter uncertainty, range from 0.05 to 0.15. The MIDI
spectrum was difficult to calibrate, and the accuracy of the
absolute photometry is not better than 30\%, although one can see
in Fig.~\ref{fig:midi} that the MIDI spectrum agrees well with the
IRAS \citep{1986A&AS...65..607O} and the ISO spectra
\citep[affected by problems of 'gluing' between different spectral
bands, as reported in][]{1999PhDT.........3V}. The log of the
observations is given in Table~\ref{tab:log_obs}. We used two
different MIDI data reduction packages: MIA developed at the
Max-Planck-Institut f\"ur Astronomie and EWS developed at the
Leiden Observatory (MIA+EWS\footnote{Available at\\
  \url{http://www.strw.leidenuniv.nl/~nevec/MIDI}},
ver.1.5.1). The resulting visibilities can be seen in
Fig.~\ref{fig:midi_vis}.


\subsection{NACO/VLT adaptive optics assisted imaging}

We observed HD\,87643 at the ESO/Paranal observatory with the NACO
adaptive optics camera \citep[NAos adaptive optics system combined with
the COnica camera,][]{2003SPIE.4839..140R}
attached to UT4 of the Very Large Telescope (VLT). NACO was operated
in the visual wavefront sensor configuration. We observed the target
with Ks ($2.2 \mu$m) and L$'$ ($3.8 \mu$m) broad-band filters. The
star HD\,296986 was used to derive the point-spread function (PSF). The
S13 camera mode was used for Ks, with a 13\,mas per pixel scale and a
14$\arcsec$$\times$14$\arcsec$ field of view. In L$'$, using 
camera mode L27, the field of view was 28$\arcsec$$\times$28$\arcsec$
and the pixel scale was 27.1~mas. The auto-jitter mode was used,
which, at each exposure, moves the telescope in a random
pattern within a box of side 7$\arcsec$ in Ks and 15$\arcsec$ in L$'$. The
journal of observations can be found in Table~\ref{tab:NACO}.

\begin{table}[htbp]
  \caption{
    Journal of observations with NACO/VLT. All stars were observed
    with a neutral density filter.
    \label{tab:NACO}
  }
  \begin{center}
    \begin{tabular}{lccccc}
      \hline
      \hline
      Date             & Star      & Used   & Used   & Exp   & Seeing\\
      &           & filt. & cam. &   time     &       \\
      \hline
      20/03/2008, 6h22 & HD\,87643   & K$_s$  & S13    & 96s        & 1.13'' \\
      20/03/2008, 6h36 & HD296986 & K$_s$  & S13    & 96s        & 1.10''  \\
      \hline
      20/03/2008, 6h55 & HD\,87643   & L$'$   & L27    & 280s$^{*}$ & 1.48'' \\
      20/03/2008, 7h21 & HD296986 & L$'$   & L27    & 324s       & 1.03'' \\
      \hline
      21/03/2008, 3h33 & HD\,87643   & K$_s$  & S13    & 96s        & 0.93''\\
      21/03/2008, 3h37 & HD\,87643   & K$_s$  & S13    & 96s        & 0.90''\\
      21/03/2008, 3h43 & HD\,87643   & K$_s$  & S13    & 84s        & 0.91''\\
      21/03/2008, 3h57 & HD296986 & K$_s$  & S13    & 96s        & 0.63''\\
      \hline
      21/03/2008, 4h21 & HD\,87643   & L$'$   & L27    & 315s       & 1.14''\\
      21/03/2008, 4h44 & HD296986 & L$'$   & L27    & 270s       & 1.73''\\
      \hline
    \end{tabular}
  \end{center}
  $^{*}$ saturated
\end{table} 

The data reduction was performed using our own scripts. First, bad
pixels were removed (i.e. interpolated using the adjacent pixel
values) and a flat-field correction was applied to the data. Then,
the sky was computed as the median of all exposures and subtracted
exposure by exposure. A visual inspection of each exposure allowed
us to check the PSF quality. The ``bad'' ones were removed for the
next step. This allowed us to significantly improve the final data
product image quality compared to the standard pipeline-reduced
frames.  This also led us to disregard the 20/03/2008
data-sets, which were in any case flagged as ``failed'' in the
observing log. A cross-correlation technique was then used to
re-centre the images with about 1 pixel accuracy. Finally, all the
selected frames were co-added, resulting in the total exposure time
shown in Table~\ref{tab:NACO}.

As a result, we got a pair of science star/PSF star images for each
observation (which were repeated due to changing weather conditions
and saturation of the detector). The PSF FWHM is 75$\pm$4~mas in the
K$_s$-band and 111$\pm$6~mas in the L'-band. In addition to a
deconvolution attempt presented in Sect.~\ref{sect:nacointerpret}, we
computed radial profiles to increase the dynamic range from about
$\approx10^3$ per pixel to $\approx10^4$-$10^5$. The result can be
found in Fig.~\ref{fig:NACO_RADIAL_PROFILES}.


\subsection{2.2m/WFI large-field imaging}

We retrieved unpublished archival ESO/Wide-Field Imager (WFI)
observations of the nebula around HD\,87643 from the ESO database
\footnote{\url{http://archive.eso.org}}
carried out on March 15 and 16, 2001. The WFI is a
mosaic camera attached to the ESO 2.2m telescope at the La Silla
observatory. It consists of eight 2k$\times$4k CCDs, forming an
8k$\times$8k array with a pixel scale of 0.238" per pixel. Hence, a
single WFI pointing covers a sky area of about 30'$\times$30'. The
observations were performed in the B, V, R$_c$ and I$_c$ broadband
filters and in the narrow-band filter in the H$\alpha$ line
($\lambda=658{\rm nm}$, ${\rm   FWHM}=7.4 {\rm nm}$). However,
we used only the observations with B, V, and R filters given that
it is a reflection nebula only \citep{1981A&A....93..285S}. In
order to cover the gaps between the WFI CCDs and to correct for
moving objects and cosmic ray hits, for each pointing, a sequence
of five offset exposures was performed in each filter.

The data reduction was performed using the package \texttt{alambic}
developed by B. Vandame based on tools available from the
multi-resolution visual model package (MVM) by
\citet{1997ExA.....7..129R}. The image is shown in
Fig.~\ref{fig:fig_WFI}.

\section{New facts about HD\,87643}
\label{sect:newfacts}



\subsection{Interferometry data: A binary star plus a compact dusty disc}
\label{sect:biarysta}

From the sparse 2006 AMBER medium spectral resolution data, the
structure of the source could not be inferred unambiguously.
The squared visibilities from 2006 did not show any significant
variation with spatial frequency. The closure phases were
measured as zero within the error estimates (see
Fig.~\ref{fig:AMBER_V2}).

The Br$\gamma$ emission line appears clearly in the AMBER
calibrated spectrum after using the technique described in
\citet{1996ApJS..107..281H} of fitting and removing the
Br$\gamma$ line for the calibrator. The resulting Br$\gamma$
line accounts for 15\% of the continuum flux at its maximum and
is spread over 5 spectral pixels (250\kms), similar to what is
reported in \citet{1988ApJ...324.1071M}.

\begin{figure}[htbp]
  \centering
  \begin{tabular}{ccc}
    \includegraphics[width=0.48\hsize, angle=0]{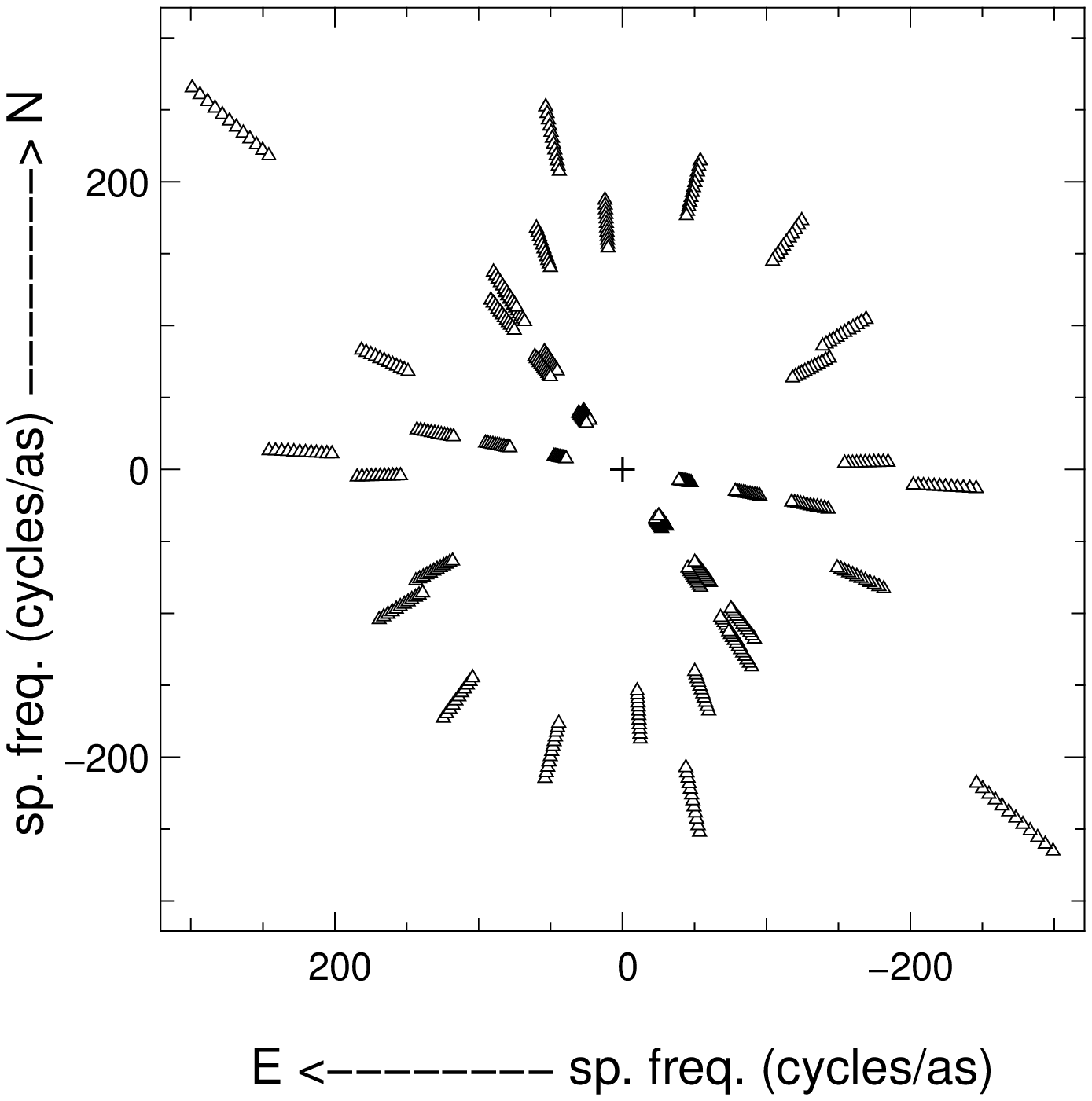}&
    \includegraphics[width=0.48\hsize, angle=0]{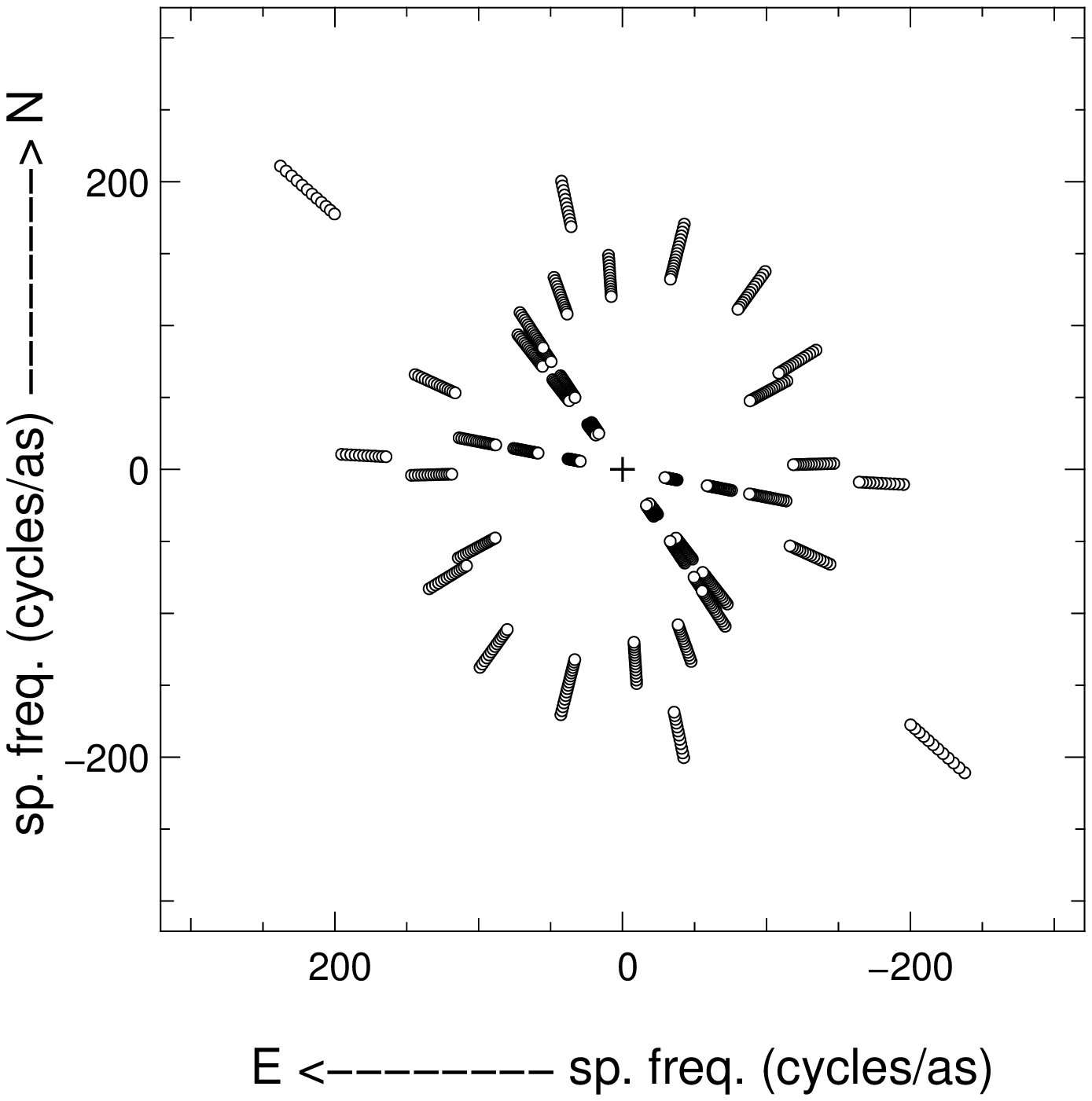}
  \end{tabular}
  \caption{
    UV coverage obtained on HD87643 in the H-band (left), in the K-band
    (right). The radial lines are caused by the different wavelengths
    within the bands.
    \label{Fig:uvCovHD87643}
  }
\end{figure}

The extensive AMBER low spectral resolution data, secured in 2008 (see
Fig.~\ref{fig:AMBER_V2}), show large variations of $V^2$ and closure
phases and clearly indicate a modulation from a binary source. Given
the good $uv$ coverage (see Fig.~\ref{Fig:uvCovHD87643}) and the
quality of the data, we undertook an image reconstruction process,
aiming to determine the best models to use for the interpretation.

\subsubsection{Image reconstruction}
\label{sect:imageReconstruct}

\begin{figure*}[htbp]
  \centering
  \includegraphics[height=0.95\textwidth, angle=-90]{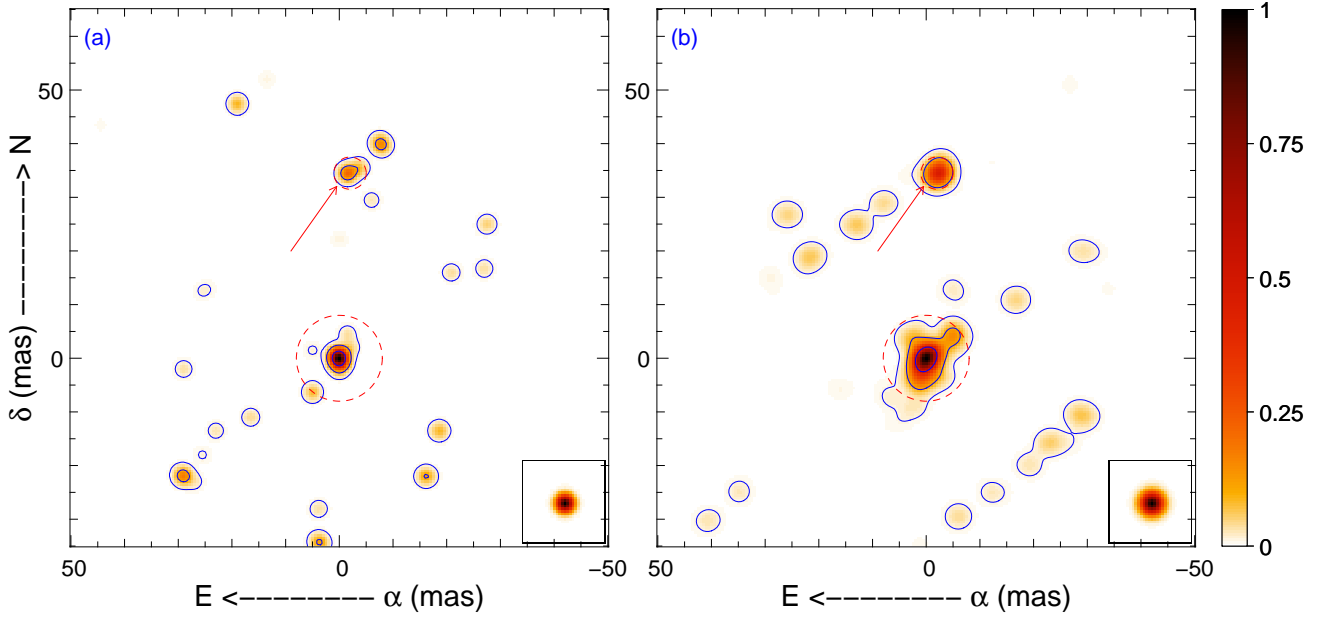}
  \caption[fig_AMBER_IMAGE]{
    \footnotesize{
      Aperture-synthesis images of HD87643 reconstructed from the
      H-band AMBER data (panel~\ref{fig:fig_AMBER_IMAGE}a) and the
      K-band data (panel~\ref{fig:fig_AMBER_IMAGE}b), assuming the
      source is achromatic in each band. Contours with 50, 10, and 1\%
      of the maximum flux are shown, and the beam size is shown in the
      lower-right box. The
      arrow shows the companion star, whereas the dashed circles show
      the trusted structures in the images.
    }
  }
  \label{fig:fig_AMBER_IMAGE}
\end{figure*}

The images in Fig.~\ref{fig:fig_AMBER_IMAGE} were reconstructed from
the AMBER squared visibilities and closure phases using the MIRA
software \citep{2008SPIE.7013E..43T}. MIRA compares the visibilities
and the closure phases from a modelled image with the observed data
using a cost-estimate optimisation, including {\it a priori}
information such as image positivity (all image pixels are positive)
and compactness of the source \citep[using a so-called {\it L2-L1}
regularisation, see][for details]{2008SPIE.7013E..43T}.

Separate image reconstructions were made using the H-band and
K-band data, assuming that the object was achromatic over the
whole band considered. MIRA needs a starting image, but we found
that the image reconstruction result did not significantly
depend on it. Extensive tests and comparisons with other
software \citep[][see
appendix~\ref{section:testsMIRA}]{1993A&A...278..328H,
  2008SPIE.7013E.121B} made us confident that we could
distinguish between artifacts and real structures in the
images. The theoretical field of view, computed using the
formula $2.44\lambda/B_{\rm min}$, is $\approx 70$\,mas, $B_{\rm
  min}$ being the shortest projected baseline (15\,m). We
convolved the K-band image with a Gaussian beam of FWHM
$\lambda/B_{\rm max} = 3.5$\,mas and the H-band one with a beam
of $\lambda/B_{\rm max} = 2.7$\,mas, $B_{\rm max}$ being the
longest projected baseline (128\,m). The image noise is of the
order of 1\%.

The reconstructed image reveals the presence of a companion star with
a plane-of-sky separation of $\approx$34.5\,mas (see
Table~\ref{table-modelparams}).  the K-band image also
reveals an extended structure detected around the main
star. We cannot assess any elongation of this resolved
structure due to the poor UV coverage in the NW direction. We
emphasise that there is still a 180$^\circ$ uncertainty due to the
overall uncertainty on the AMBER closure phase sign (and
\underline{not} due to the image reconstruction process: see
appendix~\ref{section:imageNoPhase}). We will refer in the following
to ``the northern'' and ``the southern'' components.

We also extracted relative fluxes for the different components
within the dashed circles in Fig.~\ref{fig:fig_AMBER_IMAGE} (see
Table~\ref{table-modelparams}), but one has to bear in mind that
extracting such fluxes is affected not only by large errors, but
also potentially by large systematics. Therefore, we also
performed the flux measurements using model-fitting.

\subsubsection{Model fitting}
\label{sect:modelfitting}

In Section \ref{sect:imageReconstruct}., we show that the 2008
AMBER data can be directly interpreted assuming the presence of
a binary source with a  plane-of-sky separation of $\approx$34 mas and a
P.A. of $\approx0\degr$ (see Table~\ref{table-modelparams}) and
a resolved southern component. We fitted the data using such a
model to obtain more accurate component fluxes. The images show
a series of large-scale artifacts (i.e. flux whose spatial
location is poorly constrained), accounting for a large fraction
of the total integrated flux. These artifacts are related to the
lack of data with high visibilities at small spatial
frequencies. We therefore also included an extended component to
our model, fully resolved on all AMBER baselines.

\begin{figure*}[htbp]
  \centering
  \begin{tabular}{cc}
    \psfrag{Visibility}{Squared visibility}
    \includegraphics[width=0.48\textwidth]{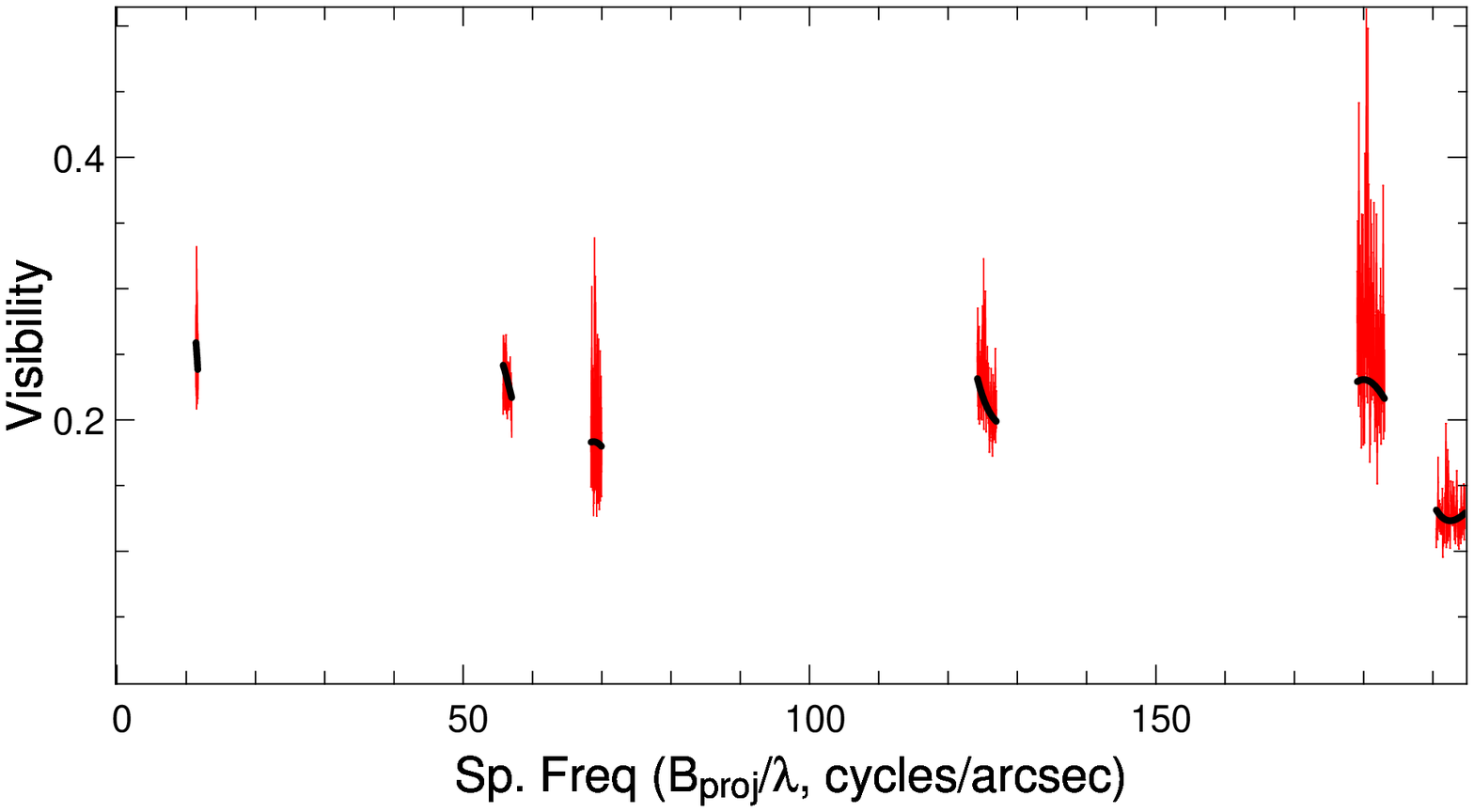}&
    \includegraphics[width=0.48\textwidth]{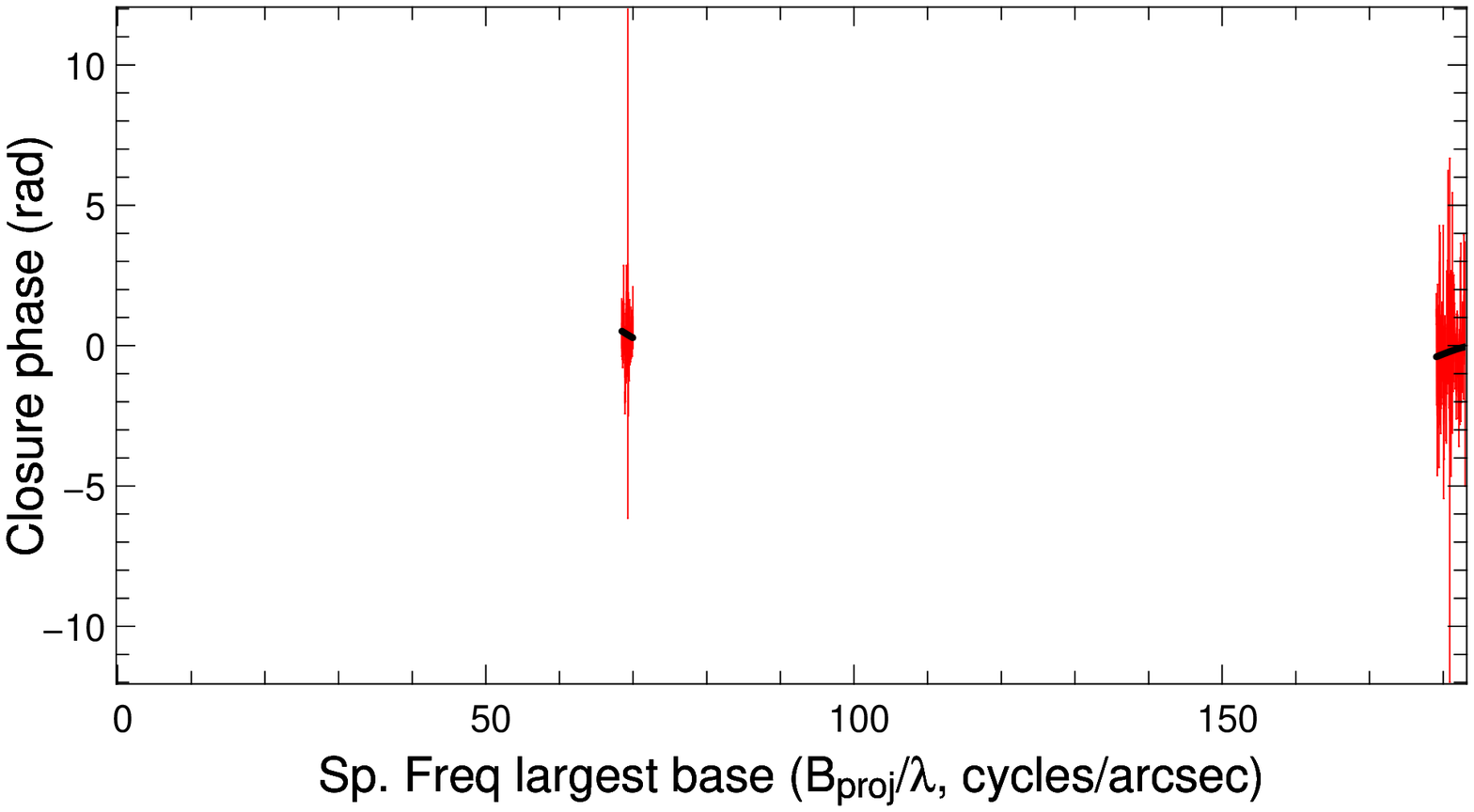}\\
    \psfrag{Visibility}{Squared visibility}
    \includegraphics[width=0.48\textwidth]{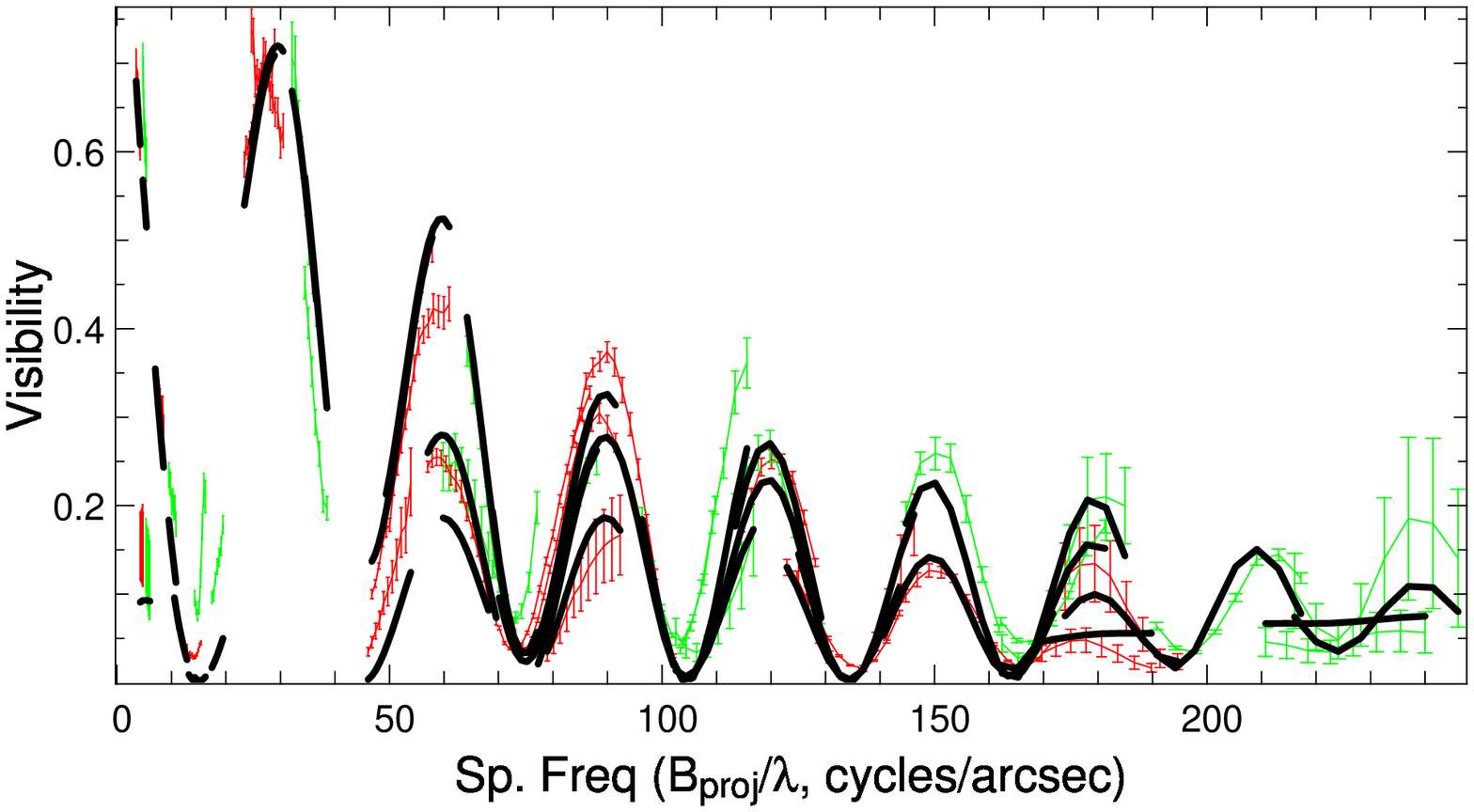}&
    \includegraphics[width=0.48\textwidth]{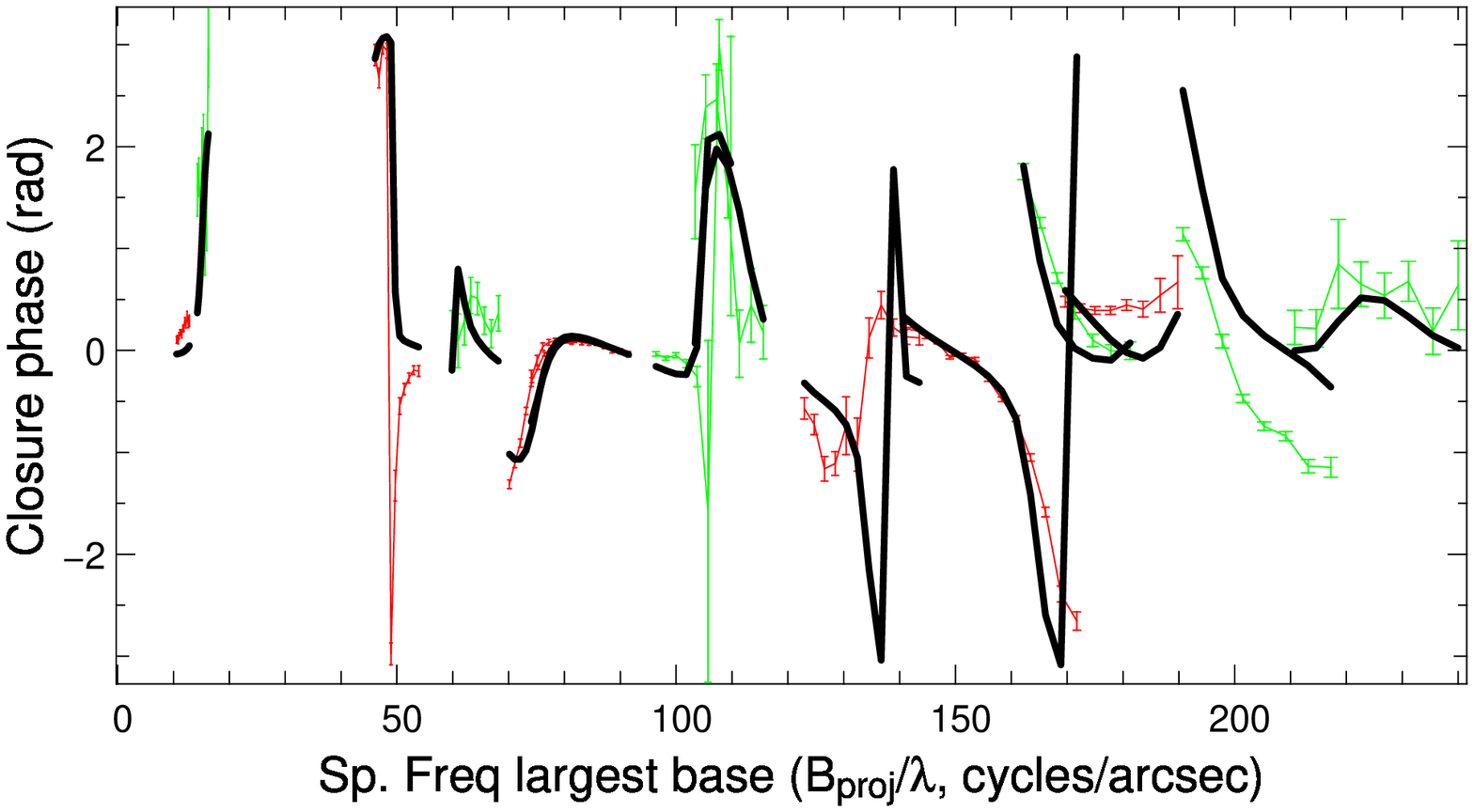}
  \end{tabular}
  \caption[fig_visibAMBER]{
    \footnotesize{
      Squared visibilities (left) and closure phases (right) plotted
      as a function of spatial frequency, projected onto the inferred
      binary position angle. The 2006 AMBER data set is seen on
      top, 2008 on the bottom. The model involving a binary system
      composed of one resolved and one unresolved component plus a
      fully resolved background is shown with a solid black line.
    }
  }
  \label{fig:AMBER_V2}
\end{figure*}

Our model-fitting used the scientific software
\texttt{yorick}\footnote{open-source scientific software freely
  available at the following URL: \url{http://yorick.sourceforge.net}}
combined with the AMBER data reduction software \texttt{amdlib}. This
tool was complemented with a series of optimisation scripts developed
by the JMMC \citep{2005sf2a.conf..271B} and others developed by
us. We judge that the resulting fits were satisfactory, even
though the formal $\chi^2$ is about 40 (see
Fig.~\ref{fig:AMBER_V2}), probably due to an underestimation of
the errors during the data reduction.

\begin{table*}
  \begin{caption}
    {
      Estimated diameter of the envelope, position of the secondary
      source, and respective fluxes, relative to the total flux. The
      envelope flux includes both the southern component and its
      compact circumstellar envelope fluxes. The errors given for the
      model-fitting reflects only formal estimates from the $\chi^2$
      value.
    }
    \label{table-modelparams}    
  \end{caption}
  \centering
  \begin{tabular}{|l|c|c|c|c|c|c|c|}
    \hline
    & \multicolumn{2}{|c|}{AMBER (model fitting)} & \multicolumn{2}{|c|}{AMBER (Image reconstruction)}& \multicolumn{3}{|c|}{MIDI (2006)}   \\
    \hline
    Wavelength (\micron) & 1.55 & 2.45 & H-band & K-band & 8.1 & 10.5  &
    12.9 \\
    \hline
    &\multicolumn{2}{|c|}{Point + Gaussian + Background}&\multicolumn{2}{|c|}{Image}& \multicolumn{3}{|c|}{Point + Gaussian + Background}\\
    \hline
    Envelope size (mas) & $3.1\pm0.1$ & $4.5\pm0.1$ & - & - &
    $\leq6.5$ & $\leq10.6$ & $\leq3.5$  \\
    \hline
    Envelope flux (\% total flux) & $61\pm2$ & $72\pm2$ & $\approx46$ & $\approx59$ &
    $17\pm2$ & $9\pm1$ & $6.3\pm0.5$ \\
    \hline
    Background flux (\% total flux) & $12\pm2$ & $9\pm2$ & $\approx44$ & $\approx25$ &
    $69\pm13$ & $77\pm9$ & $75\pm9$ \\
    \hline
    Secondary flux (\% total flux) & $27\pm1$ & $19\pm1$ & $\approx10$ & $\approx16$ &
    $14\pm3$ & $15\pm2$ & $19\pm2$  \\
    \hline
    Secondary separation (mas) & \multicolumn{2}{|c|}{$33.48\pm0.03$}
    & $34.5\pm0.5$ & $34.6\pm0.5$ &
    \multicolumn{3}{|c|}{$36.8\pm0.1$}  \\
    \hline
    Secondary PA ($^\circ$)     & \multicolumn{2}{|c|}{$-4.50\pm0.03$} & $-2.5\pm0.8$ & $-4.1\pm0.8$ &
    \multicolumn{3}{|c|}{$-26.1\pm1.7$}  \\
    \hline
  \end{tabular}
\end{table*}

\begin{figure}[htbp]
  \centering
  \psfrag{Visibility}{Squared visibility}
  \includegraphics[width=0.48\textwidth]{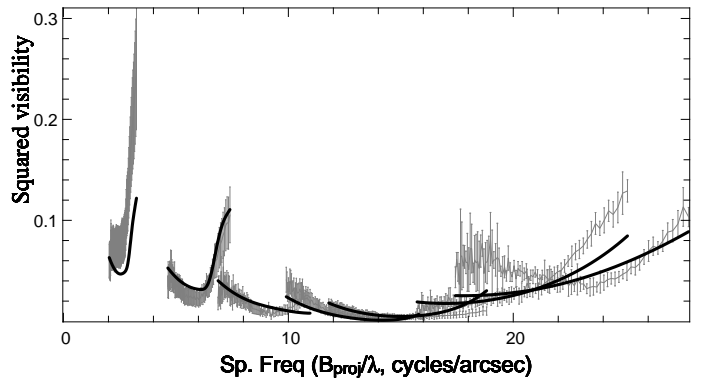}
  \caption[HD\,87643 data]{
    \footnotesize{\label{fig:midi}
      MIDI visibilities (in Gray with error bars), projected on the
      direction of the detected binary star and compared with our
      model (in thick black line).
    }
  }
  \label{fig:midi_vis}
\end{figure}

The MIDI data show complex, spectrally dependent visibility
variations from one observation to another (see
Fig.~\ref{fig:midi_vis}). Spherical and even 2-D axi-symmetric
dusty models were not able to account for this data set.  We
used the separation and P.A. inferred from the AMBER 2008 data
as initial values for the fitting
of the AMBER 2006 and MIDI 2006 measurements.

The result of such a fit shows that our binary model is
compatible with both the AMBER and MIDI 2006 data
sets. Moreover, we find that the MIDI separation is close to
that derived from the AMBER 2008 data set. The P.A. of the
secondary component differs slightly between near- and
mid-infrared, which might be due to an offset of the dust
structure compared to the main star. The total flux seen by MIDI
is dominated by a fully resolved background in addition to the
unresolved binary components
(Table~\ref{table-modelparams}). The AMBER 2006 data set is too
limited to provide a strong constraint on the the flux ratios;
therefore, we do not use them in the following analysis.

We stress that the MIDI data is affected by an independent
180$^\circ$ orientation ambiguity to the AMBER ambiguity (due to
the uncalibrated closure phase sign), as it only consists of
squared visibility data. Hence, we cannot definitely
relate the component fluxes derived from the MIDI data to those
derived from the AMBER data.

\subsection{NACO K-band imaging: a very compact dusty environment}
\label{sect:nacointerpret}

We performed a deconvolution process on our NACO images. 
Using both speckle techniques and the Lucy-Richardson deconvolution
algorithm, an elongated north-south compact structure could be
detected in the K-band but not in the L-band. This is compatible
with the AMBER data. 

\begin{figure*}[htbp]
  \centering
  \includegraphics[width=0.48\textwidth, height=0.25\textwidth, angle=0]{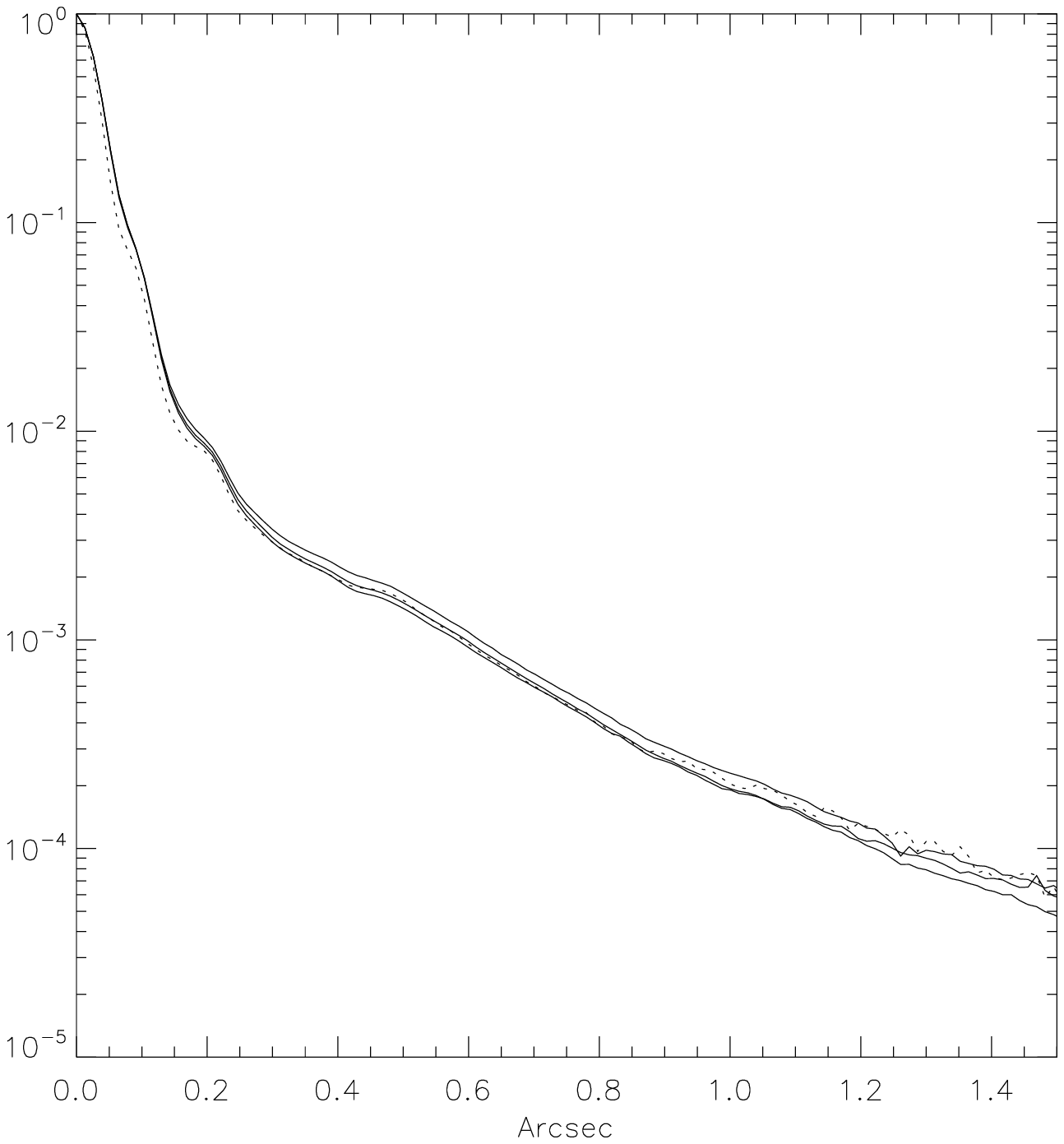}
  \includegraphics[width=0.48\textwidth, height=0.25\textwidth, angle=0]{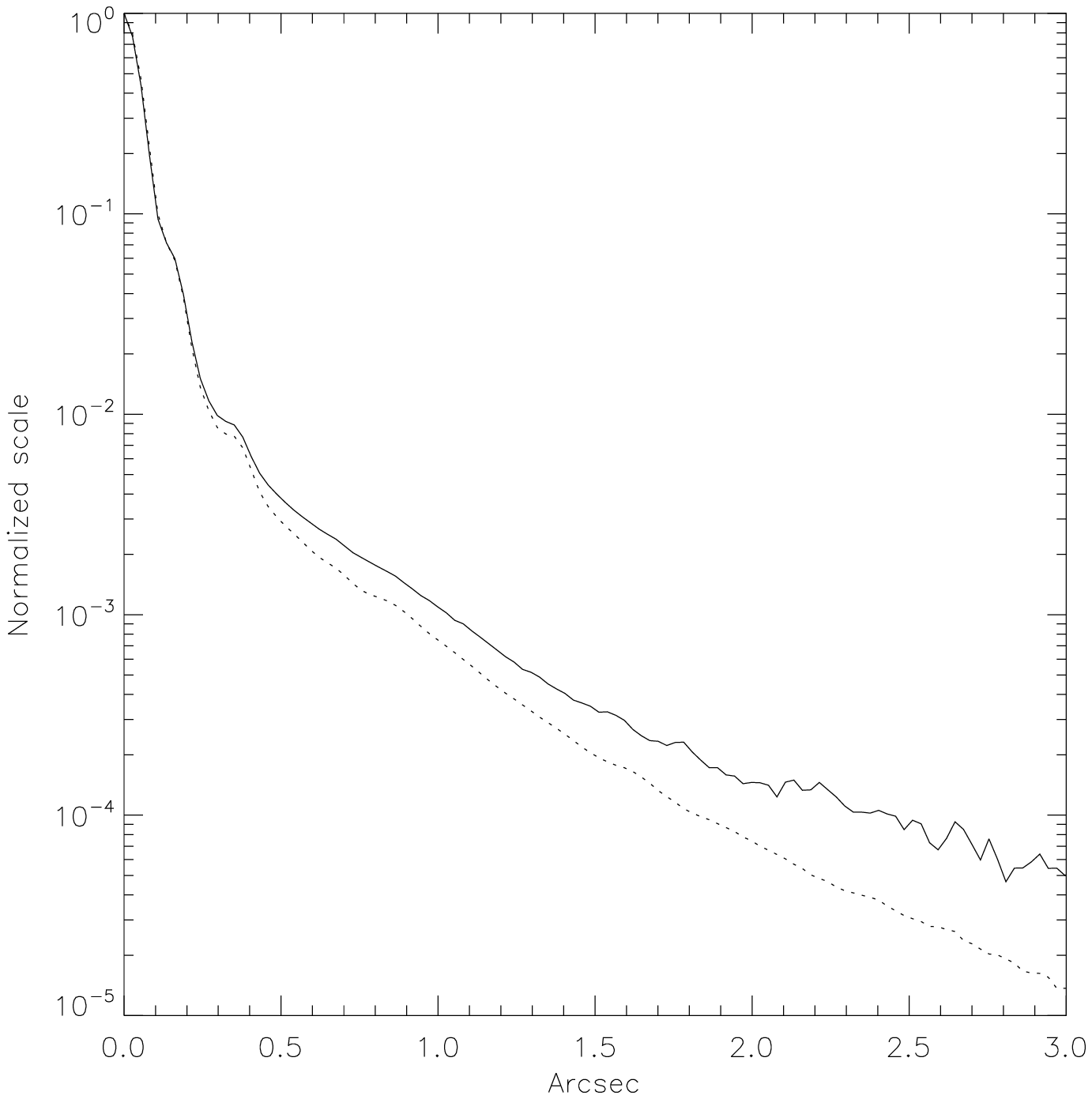}
  \caption{
    \footnotesize{
      NACO radial profiles. Left is the K$_s$ filter and right is the
      L$^{'}$ filter. HD\,87643 is shown as full lines (3
      observations in Ks, and one in L', see
      Table~\ref{tab:NACO}). The calibrator star is shown as a
      dotted line.
    }
  }
  \label{fig:NACO_RADIAL_PROFILES}
\end{figure*}

In addition, we tested the presence of dusty emission in the
surroundings of the star by azimuthally integrating the flux to
increase the dynamic range (see
Fig.~\ref{fig:NACO_RADIAL_PROFILES}). In the K-band, no extended
emission is detectable up to 1.5 arc-seconds (at larger
distances, ghost reflections or electronic ghosts impair the
dynamic range of the images) and a faint emission can be seen in
the L-band from 0.5 to 3.0 arc-seconds. This might be linked to
the fully resolved component seen in our AMBER data, and
therefore probably implies that it is much more extended than
the field of view of AMBER. However, the dynamic range of our
images ($\approx10^3$ per pixel, and $\approx10^4$-$10^5$ for the
radial profiles) does not allow us to access the spatial
distribution of such a dusty nebula.

\subsection{WFI imaging: The large scale nebula revisited}
\label{sect:wfi_neb}

The extended nebula was discovered and studied by
\citet{1972PASP...84..594V, 1981A&A....93..285S,
  1983A&A...117..359S}. In particular, using the 3.6m ESO telescope at
La Silla, \citet{1983A&A...117..359S} showed that the structure
was a reflection nebula with a 60-100" extension.

\begin{figure*}[htbp]
  \centering
  \begin{tabular}{cc}
    \includegraphics[width=0.48\textwidth]{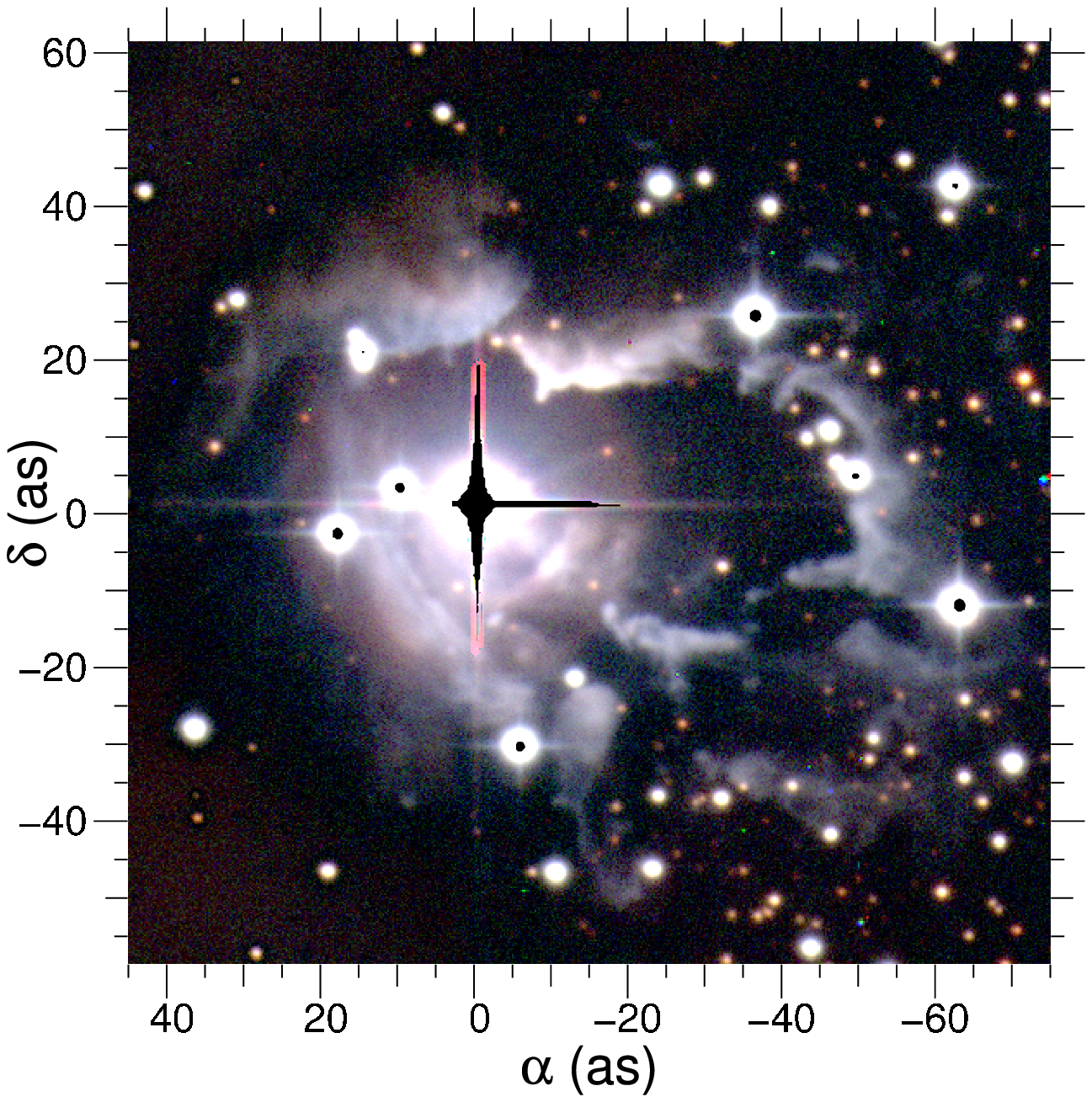}&
    \includegraphics[width=0.48\textwidth]{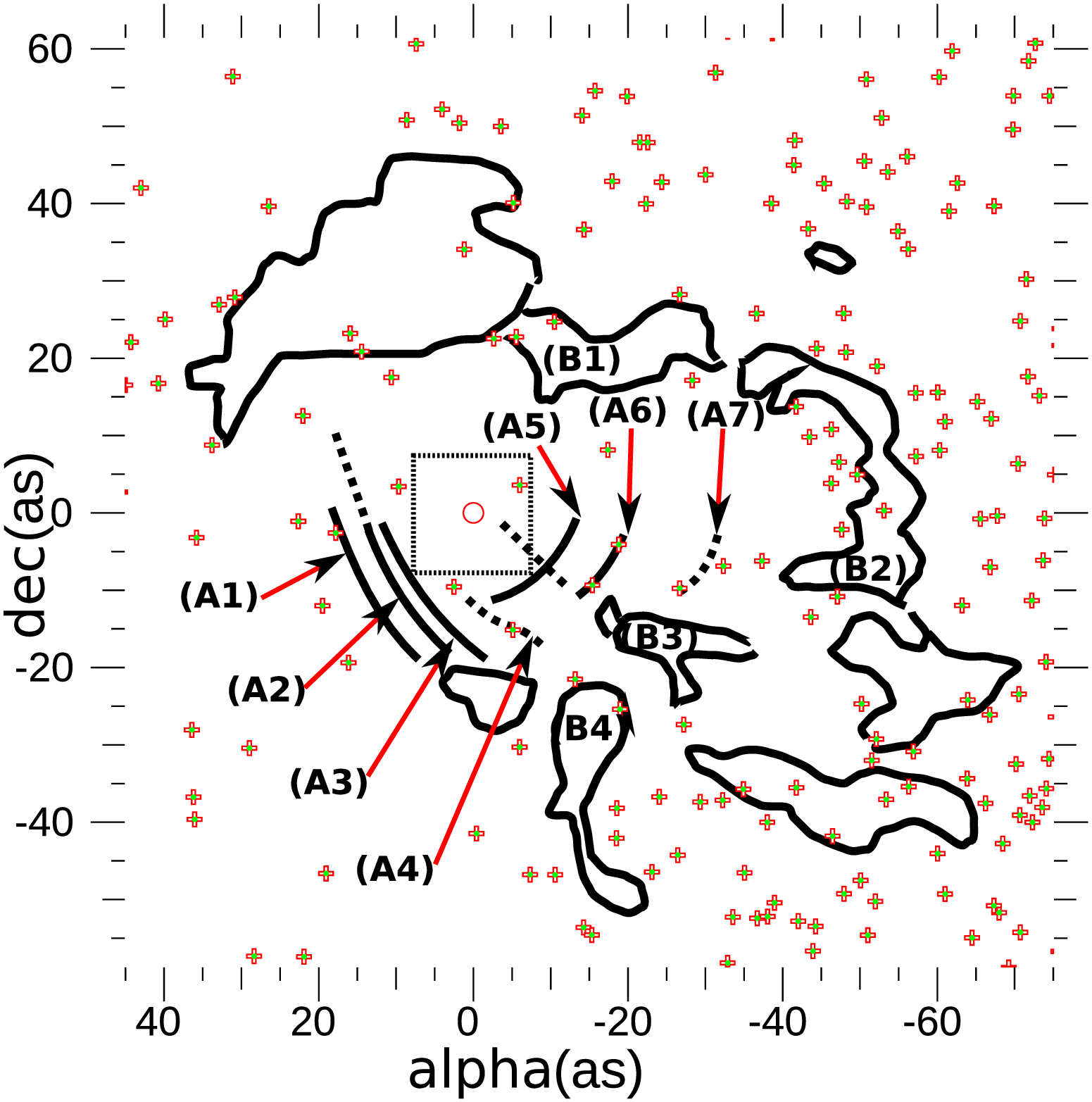}
  \end{tabular}
  \caption[fig_WFI]{
    \footnotesize{
      Reflection nebula around HD\,87643, as a composite of R, V and B
      filters (left), together with a sketch presenting the
      main structures (right). The saturated regions are masked by
      black zones in the image. This image is a small part of the
      30'x30' field-of-view of the WFI observations. The sketch shows
      HD\,87643 as a red circle, whereas other stars are marked as red
      crosses. The nebular contours are drawn as black lines and the
      prominent features are labelled (A1) to (A7) for arc-like
      structures and (B1) to (B4) for apparently blown-up nebular
      structures. Structures which are faint or uncertain are marked
      as dotted lines.
    }
  }
  \label{fig:fig_WFI}
\end{figure*}

The nebula revealed by the WFI instrument (see
Fig.~\ref{fig:fig_WFI}) with the 2.2m telescope at La Silla shows
increased dynamic range but shares the same spatial resolution and
global morphology as the previous works. The nebular structures
are primarily seen in the north-west quadrant from the central
star in the form of an extended and structured filamentary feature
at distances of 15-20'' (north) to 35-50"(west). In some
places, labelled (B1) to (B4), the nebula appears blown-up by the
central star wind, with clumpy and patchy features.

This structure has no extended counterpart in the south-east
region, where arc-like structures (labeled (A1) to (A4) in the
figure) can be seen closer to the star (at 10'', 12'', 14'', and
18'' respectively). These arcs are spaced by two to four
arc-secs and might come from past ejection events.

In the south-west region, at 12'', 19'', and 30'' respectively from the
star, two bright and one faint arc can be seen (labelled (A5) to
(A7) in the figure). These arcs do not appear to be connected to the
previously-mentioned ones.

A star count made using sextractor \citep{1996A&AS..117..393B} gives
values twice as low in the east half than in the west half of the
image. No significant relative change in the counts as a function of
filter is detectable over the field. This
indicates that there is a total absorption screen somewhere in the
field, totally masking the background stars in the east half of the
WFI image. We note that it corresponds to a dark cloud in the
  eastern part, clearly seen in our WFI image in front of red
  (hydrogen ?) nebular emission. However, we cannot directly
assess if this screen is in front or behind HD\,87643 and its
nebula.


\section{Discussion}
\label{sect:discussion}

\subsection{Adopting a distance for the system}



HD\,87643 is thought to be an evolved B[e] star, and the nebula
suggests a link to the LBV class
\citep{1983A&A...117..359S}. HD\,87643 stands out as unusual
compared to typical sgB[e] stars such as CPD-57 2874,
since ISO SWS and LWS spectra show a considerable amount of cold
dust (T$\leq$150K).

\citet{1988ApJ...324.1071M} implicitly assumed that HD\,87643 is a
supergiant in order to determine its distance, based on its
location in the direction of the Carina Arm. Such a location would
place it at a distance of 2-3\,kpc. 
However, the distance of HD\,87643 has
never been  accurately measured. When  other estimates
of the distance are found in the literature, the values are usually
much smaller. These were derived by:

\begin{enumerate}
\item \citet{1972PASP...84..594V} and \citet{1982A&A...108..111D}:
  530\,pc, based on Kurucz models for a B2V star (T$_{\rm eff}$ =
  18000~K, E(B-V) = 0.63\,mag, R$_*$ = 6 \rsun);
\item \citet{1981A&A....93..285S}: 1\,kpc, based on M$_V$
  = -4.1 derived from photometry;
\item \citet{1990ApJS...73..461S}: 1.2\,kpc, using the IUE UV
  spectrum;
\item \citet{1992A&A...261..482L}: 2-3\,kpc using the equivalent
  width (hereafter, EW) of Na{\sc i} lines. We note, however, that
  the high reported value of 2.9\,kpc is affected by a factor of 2
  error in their application of the statistical relation cited
  by \citet{1973asqu.book.....A}. Considering the correct
  relation ($r = 2.0 D$, r being the distance in kpc and D being
  the mean EW in \AA\ of the two Na D lines), the distance would be
  1.46\,kpc  (Lopes, private communication). The EW
  were measured using low-resolution spectra, implying a possible
  contamination of circumstellar origin;
\item  \citet{1998MNRAS.300..170O}: 1-6\,kpc, also based on
  the Na{\sc i} D-line equivalent widths derived from a
  higher-resolution spectrum, but using data on the interstellar
  extinction for nearby stars instead of the relationship from
  \citet{1973asqu.book.....A}. The large uncertainty of this
  distance is due to the large scatter of the extinction values in
  the Carina arm line of sight.
\item \citet{1998ASSL..233...27Z}: 1.45\,kpc, based on a
  SED fitting using various assumptions.
\end{enumerate}

Therefore, we arbitrarily use the distance of 1.5\,kpc, as suggested
by some of these works, in the subsequent parts of the present
article.

\subsection{Orbital characteristics of the binary system}

One of the main results from the AMBER interferometric observations is
that HD\,87643 is a double system. In the K-band, the NACO image core
is also significantly resolved with an elongation in the same
direction as the AMBER binary. The separation and position angle of
the components are well constrained by the data. It is more
difficult to derive an accurate flux ratio, given the complexity of the
object. Here, we shall summarise our results, from the most
constrained ones to the most speculative.

\begin{itemize}
\item The plane-of-sky separation of the components is
  $34\pm0.5$\,mas. At 1.5\,kpc, it corresponds to a
    projected separation of $51\pm0.8$\,AU. As a comparison,
  the separation of the interacting components of
  $\beta$ Lyrae \citep{2008ApJ...684L..95Z} is of the order of
  0.2\,AU (orbital period 12\,d), the separation of the
  $\gamma^2$ Velorum \citep{2007A&A...464..107M} colliding-wind
  components is of the order of 1.3\,AU (period 78\,d), and the
  approximate separation of the high-mass binary
  $\theta^1$~Ori~C \citep{2008A&A...487..323S,
    2008ApJ...674L..97P, 2007A&A...466..649K} is 15-20\,AU
  (orbital period of between 20 and 30\,yrs).
\item Therefore, the orbital period of
  HD87643 is large, approximately 20-50\,years. In
  particular, no significant orbital motion is seen  between our
  2006 and 2008 data sets.
\item The orbital plane might be seen at a high inclination
  (i.e. close to edge-on). Indeed, this is suggested
    by the high level of observed polarisation
  \citep{1998MNRAS.300..170O}. The large-scale measurements also
  seem to imply a bipolar morphology (see
  Sect.~\ref{sect:WFiimage}). In addition, HD87643 shows
  photometric variability. The short-term variations
  \citep[e.g. amplitude $\sim$0.5mag, seen in the ASAS light
  curves and in][]{1998ASSL..233..145M} are similar to
  Algol-like variability \citep{1998A&AS..131..401Y} and are
  therefore probably a consequence of the time-variable
  absorption by the material passing through the
  line-of-sight. This is also in favour of a highly inclined
  (edge-on) disc-like structure. 
\item The orbit might be highly eccentric. This hypothesis
  is supported by the periodic structures seen in the large-scale
  nebula (see Sect.~\ref{sect:WFiimage}) that would suggest
  periodic eruptions. These eruptions would be a sign of close
  encounters of the binary components, while we observed a
  well-separated binary.
\end{itemize}

\subsection{A link to the larger-scale nebula}
\label{sect:WFiimage}

Dust is found at large distances east and west of the nebula,
with fluxes reaching 15\,Jy at 16.7\micron\ (ISO/CAM1, offset of 25"
from the star, aperture 20"x14"), whereas no flux is detected in the
north and south positions\footnote{These measurements can be found
  in the ISO database: \url{http://iso.esac.esa.int}}. This points 
to a bipolar nebulosity in the east-west direction.


As mentioned by \citet{1981A&A....93..285S}, the number of field
stars decreases in the south-east direction. Also from this work
(Sect.~\ref{sect:wfi_neb}), the number of visible stars in the eastern
part of the WFI image of the nebula is about 2 times less than in the
western part. This may mean that the other side of a symmetrical bipolar
nebula remains hidden in our WFI image.

Finally, we note that inferring the geometry of the nebula is
potentially of great importance as this reflection nebula allows one
to study the central star from different viewing angles, i.e probing
different latitudes of the central star. The anisotropy of the star
flux was already noted in \citet{1981A&A....93..285S} and
\citet{1983A&A...117..359S}.


\citet{1998MNRAS.300..170O} and \citet{2006MNRAS.367..737B}
measured expansion velocities of
$\approx1000$\,\kms. Therefore one can estimate an ejection
time for the nebula: at 1.5\,kpc distance
and taking a 50" extent, one gets $\approx$355\,yrs. 

For the arc-like structures and the same value for the expansion
speed and distance, we find the following values:
$\approx$71\,yrs for A1, $\approx$85\,yrs for A2,
$\approx$100\,yrs for A3, and $\approx$128\,yrs for A4. This
first series gives ejection time intervals of $\approx$14,
$\approx$14 and $\approx$28\,yrs, respectively, between two
consecutive arcs. This might be the trace of a periodic ejection
with a period of $\approx$14\,yrs on the assumption that one arc
(between A3 and A4) is not seen in our image. However, this
periodicity is probably affected by a projection effect since
the $v\sin i$ of the arcs is not known and they appear to be
almost linear (instead of circular); hence, the $\approx$14\,yrs
periodicity is a lower limit. Concerning the second series of
arcs, we find $\approx$85\,yrs for A5, $\approx$135\,yrs for A6, 
and $\approx$213\,yrs for A7. In this case, the apparent ejection time
intervals between two consecutive arcs are $\approx$50\,yrs and
$\approx$78\,yrs. Given that the arcs appear almost circular in the image,
we can assume that the projection angle is close to
90$^\circ$. Since we have only three arcs in this case (with one
barely seen), we can only put an upper limit of $\approx$50\,yrs
on the periodicity of these ejections.

These broken structures suggest short, localised ejection that
might coincide with short periastron passages of the companion,
triggering violent mass-transfer between the
components. Therefore, we tentatively infer limits between 14
and 50\,yrs for the periodicity of the binary system.
Monitoring the system at high angular resolution over a
timescale of a few dozen years would most likely bring an
unprecedented insight into this system.

\subsection{The nature of HD\,87643}

Our interferometric measurements show a complex object composed
of a partially resolved primary component, a compact secondary
component, and a fully resolved component (i.e. extended, or
nebular, emission). As shown in Table~\ref{table-modelparams},
their relative flux strongly varies between 1.6 and 13$\mu$m.

The L-band NACO image of HD\,87643 cannot be distinguished from
the calibrator star, and there is very little emission at 1"-5"
distance. In the K-band, except from the binary signature, no
flux can be detected at a larger distance. Given the amount of
dust in the system \citep{1988ApJ...324.1071M}, the compactness
of the near-IR emissions indicate that most of it resides in a
region smaller than $\approx$100\,mas.
We may assume that the K-band
flux comes entirely from the very central source, as seen by NACO
(i.e., it contains the central binary star plus the extended
emission, as seen by AMBER). We
assume the same for the H-band. Given these hypotheses, we can infer
the absolute flux of each component using (for example) the 2MASS
magnitudes in the H and K-bands.

\begin{figure}[htbp]
  \centering
  \includegraphics[width=0.48\textwidth]{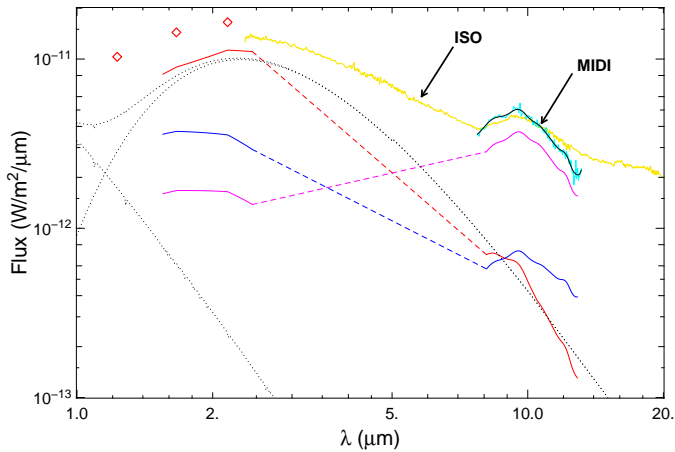}
  \caption{
    A view of the (non-de-reddened) HD87643 SED with the
    extracted fluxes from our interferometric measurements. The
    southern component flux is shown in red (top curve at 2
    \micron), the northern component flux  is shown in blue
    (middle curve at 2\micron), and the resolved background flux
    is shown in pink (bottom curve at 2 \micron). The dotted
    lines correspond respectively to a Kurucz spectrum of
    a B2{\sc V} star and to a black-body flux at 1300\,K,
    for comparison. \label{fig:sedComponents}}
\end{figure}

In the mid-infrared, the MIDI spectrum is close to the ISO
one. Therefore, we can also assume that all the N-band flux
originates from the MIDI field of view (i.e $\sim$1") and infer
the absolute fluxes of the components in the N-band. These fluxes
(N-band) and magnitudes (H and K-bands) are presented in
Table~\ref{tab:fluxes} and plotted in Fig.~\ref{fig:sedComponents}
as a function of the wavelength.

\begin{table}[htbp]
  \caption{
    Estimated fluxes for each component from the model fitting of
    Sect.~\ref{sect:modelfitting}. We used 2MASS magnitudes as the
    total magnitude in H and K-band and the MIDI fluxes in N-band.
    \label{tab:fluxes}
  }
  \centering
  \begin{tabular}{lcccccc}
    \hline
    Band or $\lambda$ & Primary & Background & Secondary \\
    (total value) & ($W/m^2/\mu m$) & ($W/m^2/\mu m$) &  ($W/m^2/\mu m$) \\
    \hline
    H (4.8\,mag)  & $8.5\times10^{-12}$ & $1.6\times10^{-12}$ & $3.7\times10^{-12}$ \\
    K (3.5\,mag)  & $1.2\times10^{-11}$ & $1.5\times10^{-12}$ & $3.1\times10^{-12}$\\
    8.1$\mu$m (92\,Jy)   & $7.2\times10^{-13}$ & $2.9\times10^{-12}$  & $5.9\times10^{-13}$ \\
    10.5$\mu$m (146\,Jy) & $3.5\times10^{-13}$ & $3.0\times10^{-12}$ & $6.0\times10^{-13}$ \\
    12.9$\mu$m (126\,Jy) & $1.4\times10^{-13}$  & $1.7\times10^{-12}$  & $4.3\times10^{-12}$ \\
    \hline
  \end{tabular}
\end{table}

Even if the effect of reddening on the H and K-band flux is not
negligible, a significant reddening fraction comes from the
circumstellar envelope. Therefore we performed this study on the
original measurements, without de-reddening.

\subsubsection{A dust-enshrouded star in the South}

The flux of the envelope around the southern component, from the
AMBER and MIDI data, can be qualitatively described by a 1300K
black body radiation (dotted line in
Fig.~\ref{fig:sedComponents}). Its Gaussian FWHM, inferred from the
model-fitting, is $\sim$4\,mas (i.e. 6\,AU at 1.5\,kpc) in the
near-infrared (H and K-bands), and is well constrained by the
AMBER 2008 data. These suggest an extended dusty
envelope around this component and clearly indicate that the dust
must be very close to the sublimation limit. Thus, the H and
K-band emission should mainly originate from the inner radius of
a dusty disc, encircling a viscous gas disc or a
2-component wind \citep{2003A&A...398..631P}. Estimating the
envelope extension with a ring instead of a Gaussian would lead
to a slightly smaller size. Thus, we can estimate that
the inner radius of the dusty disc is of the order of 2.5-3
AU (at 1.5\,kpc). 



\subsubsection{A puzzling dusty object in the North}
\label{sect:puzzlingobject}

By contrast, the northern component is unresolved (size, or
diameter, of the source $\leq2$\,mas, i.e. $\leq3$\,AU at
1.5\,kpc). Moreover, its flux can be accounted for
neither by a simple black-body at a constant temperature nor by
free-free emission. The slope of the flux variation between the
H and N-bands suggests a range of temperatures for the dust
  (from at least $\approx$300\,K to $\approx$1300\,K). No cold
dust can survive so close (T$\approx$300\,K at a radius $\leq1.5$\,AU)
to a putative luminous hot star without effective screening of the
stellar radiation. 



\subsubsection{A cold dust circumbinary envelope}

The resolved component shows a large increase of flux between H,
K, and N and also carries most of the silicate emission (see
Fig.~\ref{fig:sedComponents}). The binary system is therefore
embedded within a large oxygen-rich, dusty envelope whose shape
is not constrained by our data. This envelope must significantly
contribute to the reddening of both components.

In conclusion, we propose the following picture of the system:
two stars with dusty envelopes probably surrounded by a common
dusty envelope. One source might be a giant or a
supergiant hot star surrounded by a disc, in which we would
mainly see the inner rim corresponding to the dust sublimation
radius. The other source is surrounded by a compact dusty
envelope, and it is either not a luminous hot star or it is heavily
screened by very close circumstellar matter. In the N-band, the
circumbinary envelope contribution is not greater than
$\approx1$$\arcsec$, since the MIDI (aperture 1.2$\arcsec$) and ISO
(aperture 22$\times$14$\arcsec$) fluxes match very well, and is not
smaller than $\approx200$\,mas, as our MIDI data indicates.

\section{Concluding remarks}

Our work presents new observations of HD87643, including very
high angular resolution images. It completely changes the global
picture of this still puzzling object:

\begin{itemize}
\item The binary nature of the system has been proved, with a
  projected physical separation of approximately 51\,AU in 2008 at the
  adopted distance of 1.5\,kpc. The orbital period is most likely
  several tens of years, and the large-scale nebula indicates a
  possible high eccentricity.
\item The temperature of the southern component is
  compatible with an inner rim of a dusty disc. In the near-IR,
  we do not see the central star, but it might be a giant or a
  supergiant star.
\item The northern component and its dusty environment are
  unresolved. The underlying star is unlikely to be a massive hot
  star, or is heavily screened by close-by circumstellar material, as
  cold dust (T$\approx$300\,K) exists closer than 1.5\,AU from the
  star (if it lies at 1.5\,kpc).
\item The system is embedded in a dense, circumbinary, and dusty
  envelope, larger than 200\,mas and smaller than 1$\arcsec$.
\end{itemize}


High angular resolution observations were definitely vital in partly
revealing the nature of this highly intriguing object. 

Therefore, we call for future observations of this system, using both
high spectral resolution spectroscopy and high angular resolution
techniques, to place this interesting stellar system on
evolutionary tracks and better understand its nature.

\begin{acknowledgements}
  F. Millour and A. Meilland are funded by the Max-Planck
  Institut f\"ur Radioastronomie. M. Borges Fernandes works with
  financial support from the Centre National de la Recherche
  Scientifique (France).
  The research leading to these results has received funding from the
  European Community's Sixth Framework Programme through the Fizeau
  exchange visitors programme.
  This research has made use of services from the
  CDS\footnote{\url{http://cdsweb.u-strasbg.fr}}, from the
  Michelson Science Centre\footnote{\url{http://msc.caltech.edu}},
  and from the Jean-Marie Mariotti
  Centre\footnote{\url{http://www.jmmc.fr}} to prepare and
  interpret the observations.
  Most figures in this paper were produced with the scientific
  language yorick\footnote{\url{http://yorick.sourceforge.net}}.
  The authors thank K. Murakawa for fruitful discussions.

\end{acknowledgements}

\bibliography{biblio}
\bibliographystyle{aa}

\appendix

\section{MIRA imaging of HD87643: tests and reliability
  estimates}\label{section:testsMIRA}

This work made use of the MIRA software to reconstruct
images from the sparsely sampled AMBER interferometric data. In
this appendix, we present test results demonstrating
which structures in the reconstructed images are reliable.

\subsection{Altering the UV coverage}

The idea here is to find out if the structures remain qualitatively
the same in the final image when altering the resolution (removing the
high spatial frequencies) or the low frequency components of the UV
plane.

\subsubsection{Removing low spatial frequencies}

\begin{figure}[htbp]
  \centering
  \begin{tabular}{cc}
    \includegraphics[height=0.5\hsize, angle=-90, origin=c]{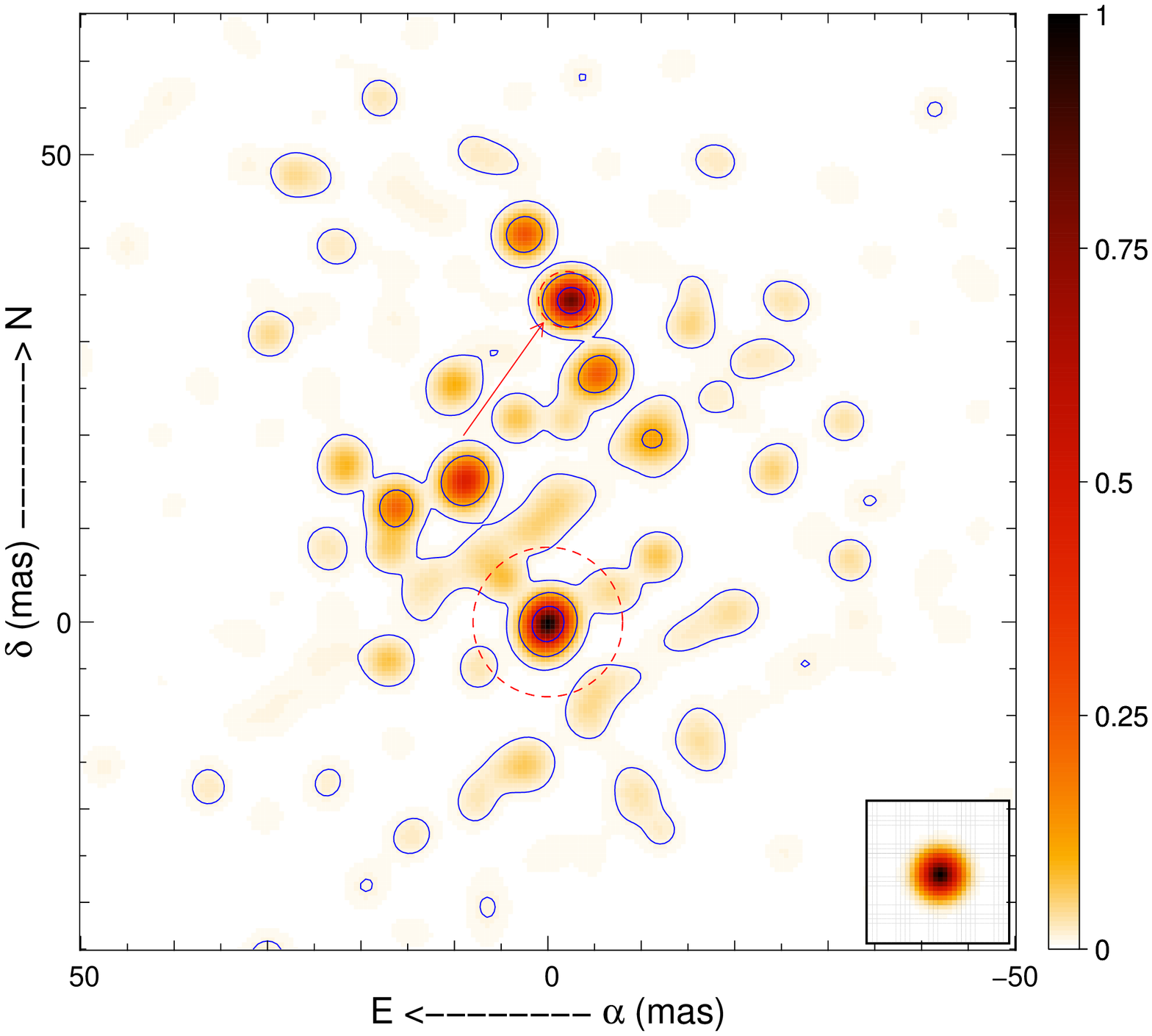}&
    \includegraphics[width=0.47\hsize, angle=-0, origin=c]{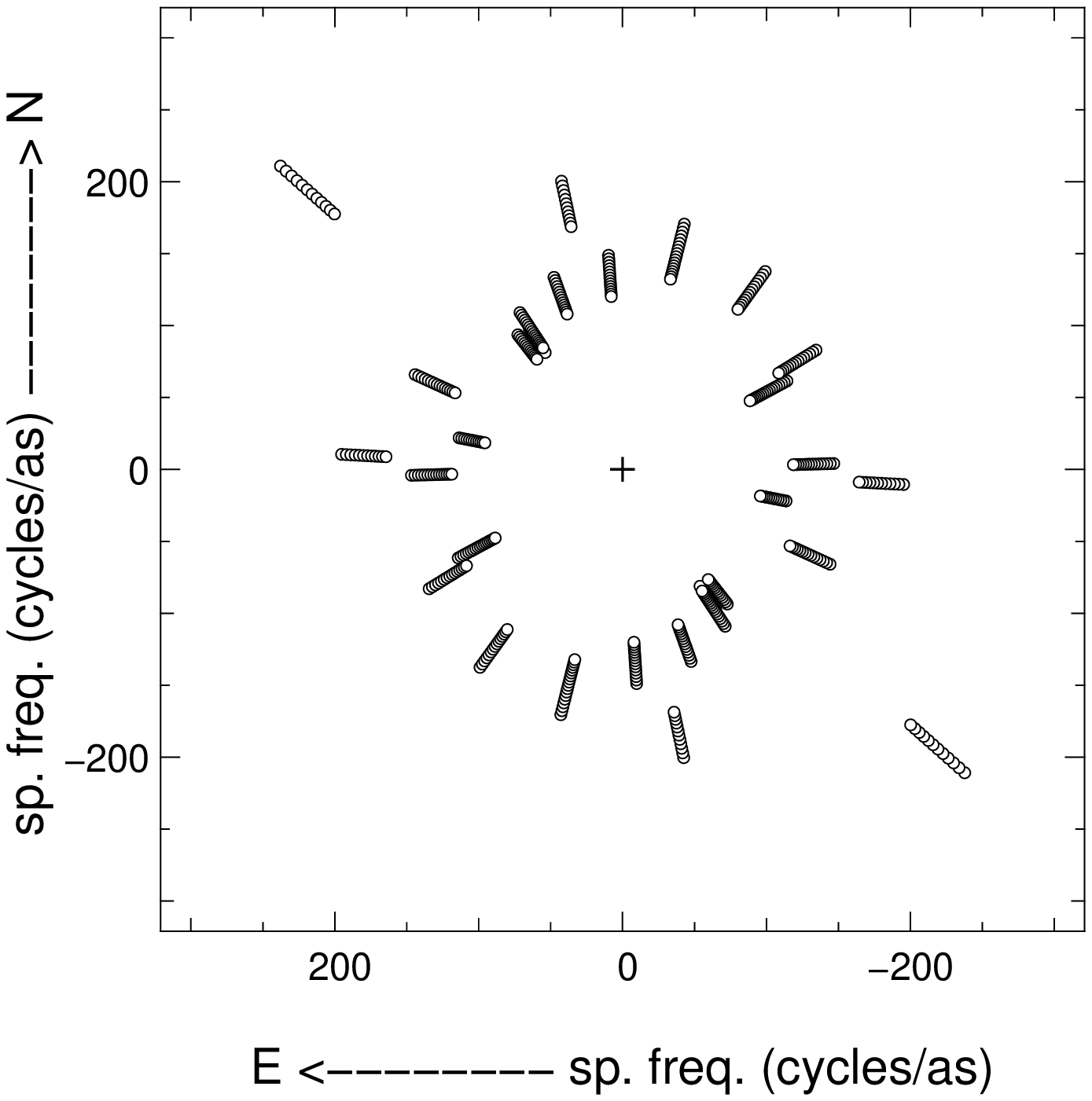}\\
  \end{tabular}
  \caption{
    Removing low spatial frequencies to check whether the structures
    are reconstructed.
    \label{Fig:cutsmallfreq}
  }
\end{figure}

Removing low spatial frequencies reduces the field of view, and
the MIRA image reconstruction software does not manage to overcome
this difficulty (see Fig.~\ref{Fig:cutsmallfreq}). Therefore, we
conclude that the low spatial frequencies (i.e. the short baselines)
are just as important as longer baselines to perform image
reconstruction.

\subsubsection{Removing high spatial frequencies}

\begin{figure}[htbp]
  \centering
  \begin{tabular}{cc}
    \includegraphics[height=0.5\hsize, angle=-90, origin=c]{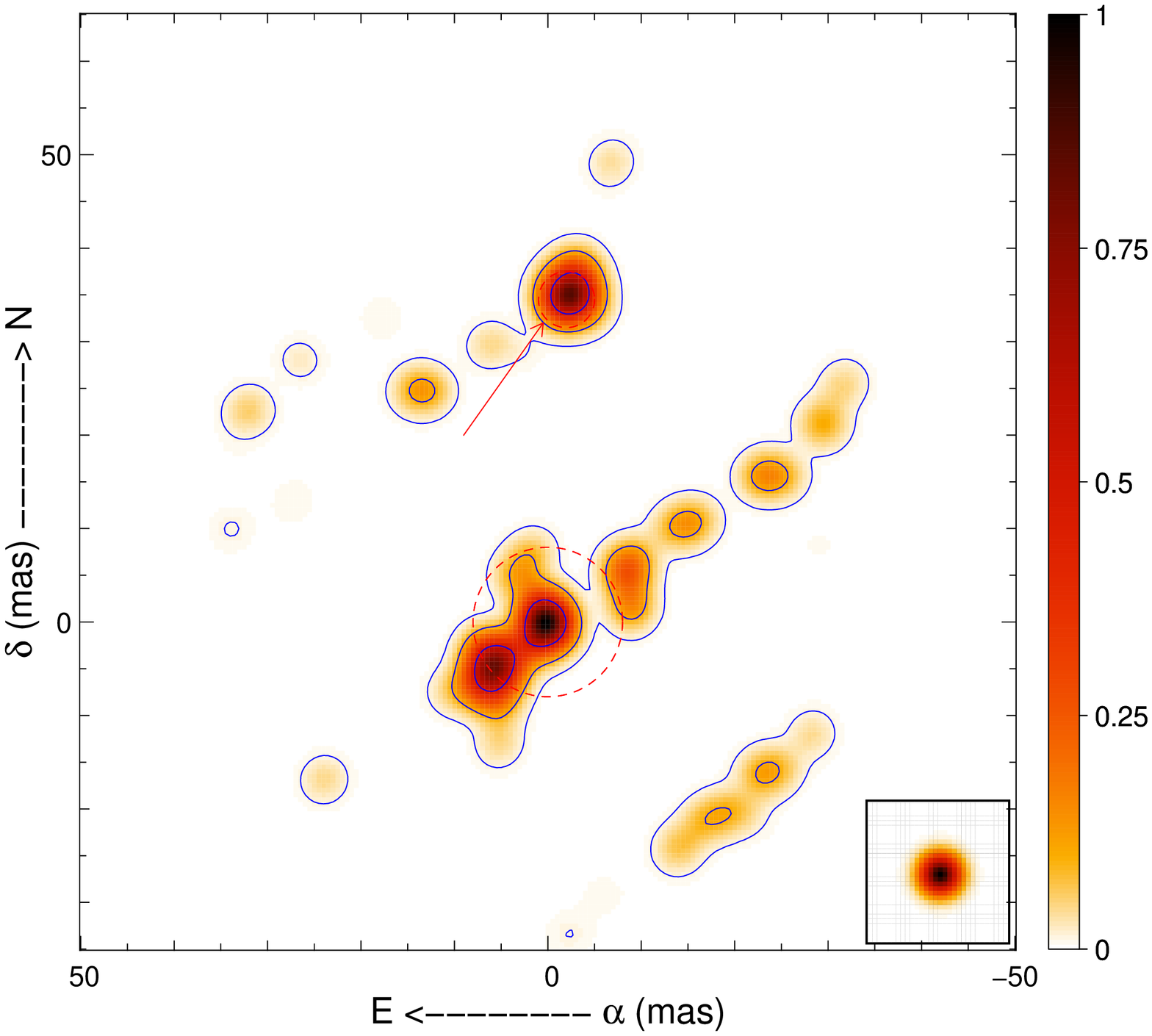}&
    \includegraphics[width=0.47\hsize, angle=-0, origin=c]{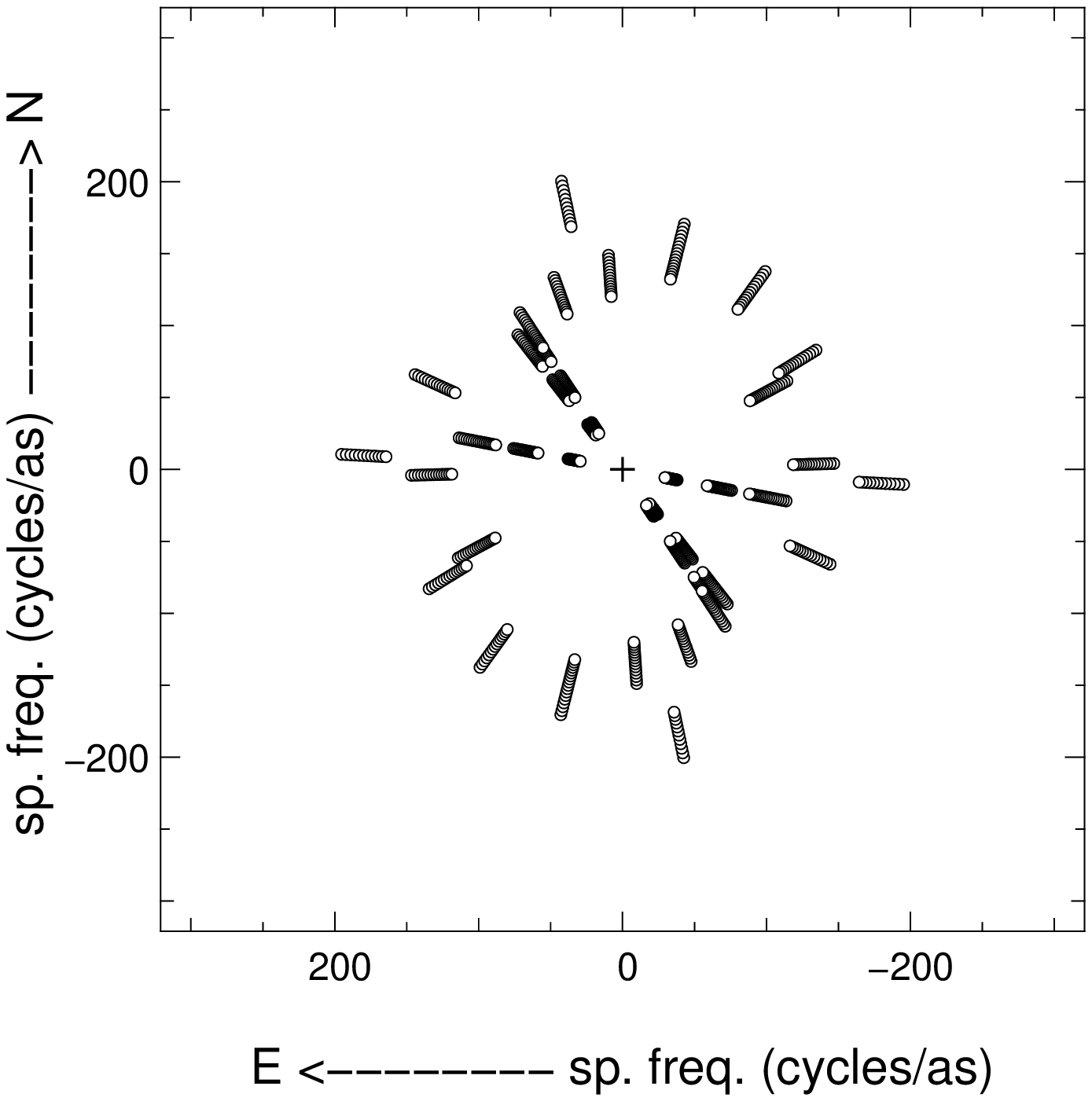}\\
  \end{tabular}
  \caption{
    Removing high spatial frequencies to have a more symmetric UV
    coverage. \label{Fig:cutLargeFreq}
  }
\end{figure}

As a second test of the image reconstruction reliability, we cut
all the high spatial frequencies in order to get a more symmetric
UV coverage. The result is seen in Fig.~\ref{Fig:cutLargeFreq},
together with the corresponding UV coverage. The striking point
compared to the image presented in Fig.~\ref{fig:fig_AMBER_IMAGE}
is that all the structures seen in this new image are
qualitatively the same as before. This ensures that the binarity
and resolution of the southern components are not due to
image reconstruction artifacts.

\subsubsection{Imaging without phases}
\label{section:imageNoPhase}


Since the AMBER closure phases in our data-set were noisy
($\sigma\approx0.5-1$\,radian), we wondered whether these phases were
contributing significant information to the image reconstruction
compared to squared visibilities alone.

\begin{figure}[htbp]
  \centering
  \begin{tabular}{ccc}
    \includegraphics[width=0.5\hsize, angle=-90]{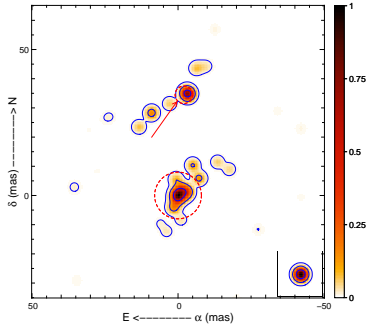}
  \end{tabular}
  \caption{
    HD87643 image reconstruction without the phase information. We
    had to rotate the image by 180$^{\circ}$ to make it match the
    previous reconstructions.
    \label{Fig:HD87643NoPhase}
  }
\end{figure}

We tried to reconstruct an image of HD87643 using the same AMBER
data, except for the closure phases, and both the MIRA and BSMEM
image reconstruction software. Indeed, they are able to cope with
phase-less data, in this case making a phase-retrieval image
reconstruction. In Fig~\ref{Fig:HD87643NoPhase} we show the MIRA
reconstruction. We are able to reconstruct qualitatively the same
image of HD87643 as in Fig.~\ref{fig:fig_AMBER_IMAGE} without
using the closure phase information.

We note, however, that we had to rotate the result  to match the
orientation of Fig.~\ref{fig:fig_AMBER_IMAGE}. This makes
sense since only the phase information would be able to orient the
system.

\subsection{Testing with other image reconstruction algorithms}

MIRA is only one example of an image reconstruction software
package for optical interferometry. Other software packages
exist, and we present here the tests we made using several of
them.

\subsubsection{BSMEM}

\begin{figure}[htbp]
  \centering
  \begin{tabular}{cc}
    \includegraphics[height=0.5\hsize, angle=-90]{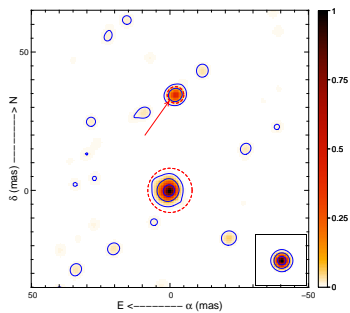}
  \end{tabular}
  \caption{
    Image reconstruction of HD87643 using the BSMEM software.
    \label{Fig:bsmemReconstruct}
  }
\end{figure}

BSMEM is based on the Maximum Entropy Method (MEM) applied to
bispectrum measurements (deduced from the squared visibilities and
closure phase measurements). It works using an iterative algorithm,
comparing the distance from the reconstructed image to the data and
the ``entropy'' computed from the properties of the image itself
\citep{2008SPIE.7013E.121B}. We applied BSMEM to the AMBER
data. The result can be seen in
Fig.~\ref{Fig:bsmemReconstruct}. We find the same structures as for
the MIRA reconstruction process.

\subsubsection{Building-Block Method}

\begin{figure}[htbp]
  \centering
  \begin{tabular}{cc}
    \includegraphics[height=0.5\hsize, angle=-90]{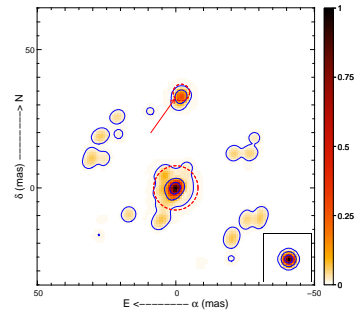}
  \end{tabular}
  \caption{
    Image reconstruction of HD87643 using the Building-Block
    software.
    \label{Fig:BBReconstruct}
  }
\end{figure}

The Building-Block method \citep{1993A&A...278..328H} was
developed to reconstruct diffraction-limited images from the
bispectrum of the object obtained with bispectrum speckle
interferometry and long-baseline interferometry. Since the
intensity distribution of an object can be described as a sum of
many small components, the Building-Block algorithm iteratively
reconstructs images by adding \textit{building blocks}
(e.g. $\delta$-functions). The initial model image may simply
consist of a single $\delta$ peak. Within each iteration step,
the next building block is positioned at the particular
coordinate, which leads to a new model image that minimises the
deviations ($\chi^2$) between the model bispectrum and the
measured object bispectrum elements. An approximation of the
$\chi^2$-function was derived which allows fast calculation of a
large number of iteration steps. Adding both positive and
negative building blocks, taking into account the positivity
constraint, and adding more than one building block per
iteration step improves the resulting reconstruction and the
convergence of the algorithm. The final image is obtained by
convolving the Building-Block reconstruction with a beam
matching the maximum angular resolution of the
interferometer. We applied this method to the AMBER data, and
the resulting image (Fig.~\ref{Fig:BBReconstruct}) shows many
similarities with the MIRA one, including a resolved southern
component in the K-band.

\subsection{Conclusion}

The tests presented above show that the binary and resolved
southern component are structures that can be trusted in the
images. All other structures (possible elongation of the
southern component, inclined large ``stripes'' in the images)
are artifacts from the image reconstruction process, which may
indicate that the image has some additional flux, not
constrained by the observations (fully resolved background).

\end{document}